\documentclass[iop,twocolappendix]{emulateapj}
\usepackage{amsmath,mathtools,mathrsfs}
\newcommand{\sgra}{Sgr~A$^*$}

\begin{document}
\title{Principal Component Analysis as a Tool for Characterizing Black Hole Images and Variability}
\author{Lia Medeiros\altaffilmark{1,2}, Tod R.\ Lauer\altaffilmark{3}, Dimitrios Psaltis\altaffilmark{1}, Feryal \"Ozel\altaffilmark{1}}

\altaffiltext{1}{Steward Observatory and Department of Astronomy, University of Arizona, 933 N. Cherry Ave., Tucson, AZ 85721}
\altaffiltext{2}{Department of Physics, Broida Hall, University of California Santa Barbara, Santa Barbara, CA 93106}
\altaffiltext{3}{National Optical Astronomy Observatory, Tucson, AZ 85726}

\begin{abstract}
We explore the use of principal component analysis (PCA) to characterize high-fidelity simulations and interferometric observations of the millimeter emission that originates near the horizons of accreting black holes.  We show mathematically that the Fourier transforms of eigenimages derived from PCA applied to an ensemble of images in the spatial-domain are identical to the eigenvectors of PCA applied to the ensemble of the Fourier transforms of the images, which suggests that this approach may be applied to modeling the sparse interferometric Fourier-visibilities produced by an array such as the Event Horizon Telescope (EHT). We also show that the simulations in the spatial domain themselves can be compactly represented with a PCA-derived basis of eigenimages allowing for detailed comparisons between variable observations and time-dependent models, as well as for detection of outliers or rare events within a time series of images. Furthermore, we demonstrate that the spectrum of PCA eigenvalues is a diagnostic of the power spectrum of the structure and, hence, of the underlying physical processes in the simulated and observed images.  
\end{abstract}

 
\section{Introduction}\label{sec:intro}

The task of imaging and modeling the millimeter emission close to the horizon of an accreting black hole with the Event Horizon Telescope (EHT) encompasses a number of challenges. Interferometric imaging requires accurate synthesis of an image based on a sparse and incomplete set of Fourier visibilities (see, e.g.,~\citealt{Honma2014,Bouman2016,Chael2016,Akiyama2017}). Understanding the morphological diversity of the structure of the emission and its dependence on the physical parameters of the black hole rests on the comparison of such observations to high-fidelity simulations of the accretion flow (see, e.g.,~\citealt{Moscibrodzka2009, Moscibrodzka2017, Dexter2009,2015ApJ...799....1C, 2016ApJ...832..156K,Gold2017}). The accretion flow itself is dynamic, potentially causing strong variations in the emission morphology over the very time scales required to synthesize an image with a large baseline interferometer (see, e.g.,~\citealt{2016ApJ...817..173L, 2017ApJ...844...35M, Medeiros2018}). In considering all of these issues, a common thread that emerges is a need to efficiently capture and characterize a complex series of images in diverse contexts. In this and forthcoming papers, we apply Principal Component Analysis (PCA) to General Relativistic magnetohydrodynamic (GRMHD) simulations and simulated EHT observations to explore its utility as an approach to addressing these challenges.

PCA is a mathematical approach to quantifying variability of an ensemble. In our case, the ensemble is a collection of images obtained from time-dependent simulation outputs of black hole accretion flows. PCA is non-parametric and does not incorporate any physical knowledge of the black hole or its accretion physics. Instead, PCA decomposes each image into a sum of orthogonal-basis eigenvectors (i.e., eigenimages) with eigenvalues that correspond to the brightness variance that each eigenimage captures. The eigenimages are then ranked by their eigenvalues, which allows minor variations to be discarded if desired. In other words, PCA allows for a compact and effective representation of the images in the ensemble. In practice, the implicit compression can be substantial, using perhaps only a dozen eigenvectors to represent over 1000 source images \citep{2010AJ....140..390B}. 

In this initial exploration, we show that PCA is particularly useful to help recognize and characterize the large-scale temporal variability in the morphology of the millimeter emission close to the horizon of black holes such as \sgra and M87. Numerous observations and studies in the past few decades have established the fact that black hole accretion flows are highly variable. X-ray observations of galactic black hole binaries reveal a variability spectrum characterized by red noise as well as distinct high-frequency quasi-periodic components (see, e.g., \citealt{2000ARA&A..38..717V,  2006ARA&A..44...49R}). Similarly, multi wavelength observations of nearby AGN, including the Galactic Center black hole \sgra, show variability on timescales ranging from hours to months (see, e.g., \citealt{2003Natur.425..934G, 2009ApJ...691.1021D, 2015ApJ...799..199N}). This variability is not surprising. It is understood to be the result of the turbulent accretion flows as well as a potential manifestation of the unique black-hole spacetime near its horizon (see, e.g.,~\citealt{1994ApJ...421...46R,chan2015}). Because of this, it is expected that the images for \sgra and M87 at mm wavelengths will also be variable at dynamical timescales at the vicinities of the black-hole horizons~(\citealt{2017ApJ...844...35M,Medeiros2018}). 

Black hole variability is an important consideration for the EHT. Because interferometry relies on the rotation of the Earth to obtain images, it is critical to understand how the sources may vary over timescales comparable to those observations in order to design and implement proper image reconstruction algorithms~(see, e.g., \citealt{2016ApJ...817..173L,2017ApJ...850..172J, 2017arXiv171101357B} for early attempts). It is also important to study and characterize the variability predicted by different accretion flow models, in order to investigate whether the mode and amplitude of predicted variability agrees with observations (see~\citealt{2016ApJ...832..156K}). Both these issues are especially true for the primary target for the EHT, \sgra, for which the characteristic time of variability can be as short as a few minutes. 

In this paper we show that PCA can generate a compact orthogonal set of basis eigenimages that can represent accurately the ensemble of images generated in a suite of high-fidelity GRMHD simulations and facilitate the efficient comparison of models to observations. This basis may also be used to provide a compact rendition of the ensemble in Fourier space and, in turn, a path to efficient representation of the sparse visibility observations. Furthermore, we show that PCA allows us to recognize ``outliers'' in the typical source morphology and identify, both in simulations and in observations, instances of episodic physical phenomena, such as magnetic reconnection and flaring events. 

In parallel to the efforts to characterize and understand its origins, the EHT has a nearly orthogonal interest in the question of black hole variability, i.e., identifying emission signatures that are, in fact, not variable. In particular, the black hole shadow, which offers excellent opportunities for testing the predictions of general relativity \citep{2015ApJ...814..115P}, is expected both to be invariant in time and nearly independent of any property of the system other than the mass of the black hole. It is, therefore, valuable to separate time-variable aspects of black hole images, such as turbulence and periodic variabilities in the accretion flow, from the constant signals arising from the black hole spacetime. 

The structure of this paper is as follows. We provide a brief development of the PCA formalism in $\S2$ and demonstrate that a PCA basis that is derived in the image domain also provides a basis in the Fourier (i.e., visibility) domain. We apply our formalism to a simple ensemble of images in \S\ref{sec:PCA3}. In \S\ref{sec:PCAsims}, we demonstrate the ability of PCA to represent a temporal sequence of high-fidelity simulated images of an accreting black hole. We demonstrate the use of PCA to compactify the space of images using dimensionality reduction and to identify times of rare or unusual activity in the simulated time series in  \S\ref{sec:reconstruction}. In \S\ref{sec:results3}, we compare the spectrum of PCA eigenvalues to that of Gaussian and red-noise processes and show how the PCA eigenvalues are related to the underlying power spectrum of structures in the images. We conclude and discuss future applications of our work in \S \ref{sec:conc}. 

\section{Principal Component Analysis}\label{sec:PCA}

Our goal is to use PCA to determine the dominant components in a set of images of black holes. In this section, we give a brief introduction to PCA and show that it may be applied directly to interferometric observables. The majority of this derivation follows \citet{turk1991} with some differences that we will explicitly outline below.

\subsection{Introduction to Principal Component Analysis}\label{sec:PCA1}

The principle of PCA is to calculate a set of orthogonal eigenimages (or eigenvectors) from an ensemble of images. We can then utilize this basis to compactly represent all of the images in the original ensemble as a linear combination of those eigenimages. 

We denote an ensemble of $m$ images by $I_n(x,y)$, where $n=1,...,m$ and the pair of coordinates $(x,y)$ are used to represent the location of each of the $N\times N$ pixels on the image. For simplicity, each image can also be represented as a column vector $\textbf{I}_n$ of length $N^2$. For our purposes, the ensemble of images will be comprised of a series of snapshots of a black hole accretion flow that are obtained from simulations or observations, although the derivations we provide below are much more general.

As the basis of our decomposition, we choose to use the $m$ orthogonal eigenimages $\textbf{u}_k$ of the covariance matrix 
\begin{equation}
\begin{split}
C &= \frac{1}{m} \sum_{n=1}^m  \textbf{I}_n \, \textbf{I}_n^T \\
&\equiv AA^T\;.
\end{split}
\end{equation}
In this equation, we defined the $N^2 \times m$ matrix $A$, such that its columns are the $m$ images of the ensemble, i.e.,
\begin{equation}
A \equiv [\textbf{I}_1 \,\, \textbf{I}_2 \,\, \cdots \,\, \textbf{I}_m].
\end{equation}
Strictly speaking and contrary to the notation of \citet{turk1991}, $C$ is not a covariance matrix because we have not subtracted the mean from each image. However, we will refer to $C$ as the covariance matrix throughout the paper to avoid introducing unnecessary terminology.

The covariance matrix $C$ is an $N^2 \times N^2$ matrix that measures how the variation in the brightness of each pixel across the ensemble of images is correlated to the variation in brightness of every other pixel. We can write explicitly each element of the matrix $C$ as 
\begin{equation}
C_{ij} = \frac{1}{m} \sum^m_{n=1} A_{in} A_{jn},
\end{equation}
where the indices $i$ and $j$ correspond to the $N^2$ pixels ($i,j = 1,2,...,N^2$) and the index $n=1,2,...,m$ corresponds to the different images in the ensemble.  

In principle, we can then find the eigenimages $\textbf{u}_k$ of the covariance matrix $C$ by diagonalizing it such that
\begin{equation}
\begin{split}
C \textbf{u}_{k} &= \mu_{k}\textbf{u}_{k} \\
AA^T \textbf{u}_{k} &= \mu_{k}\textbf{u}_{k}\;.
\end{split}
\end{equation}
However, diagonalizing an $N^2 \times N^2$ matrix is computationally expensive and, in fact, not necessary. Because there are (at most) only $m$ independent images in the ensemble, there are only $m$ non-trivial eigenvalues and eigenvectors for this covariance matrix, which we can compute in an efficient way.\footnote{See, e.g., Appendix A of \citet{book_lin_alg} for a discussion of this property.}

We start by computing the eigenvectors and eigenvalues of the $m\times m$ matrix $L = A^TA$ such that
\begin{equation}\label{eq:L_eigen}
\begin{split}
L \textbf{v}_\gamma &= \lambda_\gamma\textbf{v}_{\gamma} \\
A^TA \textbf{v}_{\gamma} &= \lambda_{\gamma}\textbf{v}_{\gamma}\;,
\end{split}
\end{equation}
where $\textbf{v}_{\gamma}$ are the $m$ eigenvectors of $L$, each of dimension $m$. It is then easy to show by multiplying both sides of equation (\ref{eq:L_eigen}) by $A$ that the matrix $L$ and the covariance matrix $C$ share the same eigenvalues, i.e.,
\begin{equation} \label{eq:AATA_CA}
\begin{split}
AA^T A \textbf{v}_{\gamma} &= \lambda_{\gamma} A\textbf{v}_{\gamma} \\
C A \textbf{v}_{\gamma} &= \lambda_{\gamma} A\textbf{v}_{\gamma}.
\end{split}
\end{equation}
This equation also demonstrates that the vectors 
\begin{equation}\label{eq:eigenimages}
\textbf{u}_\gamma=A\textbf{v}_{\gamma},
\end{equation}
of size $N^2$, are the eigenimages of the covariance matrix $C$ with corresponding eigenvalues $\lambda_{\gamma}$.

The normalization of the eigenimages is, in principle, arbitrary. Following standard PCA convention, we normalize each eigenimages such that 
\begin{equation}\label{eq:eigenv}
\textbf{u}_k^2=\lambda_k
\end{equation}
and
\begin{equation}\label{eq:suml}
\sum_{k=1}^m \lambda_k=1\;.
\end{equation}
Because the eigenvectors are orthogonal, it also follows that $\textbf{u}_k \textbf{u}_{k'} =\lambda_k \delta_{k k\prime}$, where $\delta_{k k\prime}$
is the Kronecker delta. Hereafter, we will use the notation $\textbf{u}_k^2=\textbf{u}_k \textbf{u}^T_k$ to denote the square of the magnitude of an eigenimage. The overall sign of each eigenimage is arbitrary and, in principle, eigenimages may contain negative fluxes. In the latter case, we further multiply the brightness of each pixel in the eigenimage by $-1$ in order to enforce the total flux of the eigenimage to be positive.

Having obtained the eigenimages of our ensemble, we can then express any of its images as the linear combination
\begin{equation}\label{eq:I_n_recon}
\textbf{I}_n = \sum_{k=1}^ma_{nk}\textbf{u}_k,
\end{equation}
where 
\begin{equation}\label{eq:amp}
a_{nk} \equiv  \frac{\textbf{I}_n \textbf{u}^T_k }{( \textbf{u}_k^2)}
\end{equation}
are the amplitudes of the projections of the images on the eigenimage basis. 

The square of the magnitude of each image is equal to
\begin{equation}
\begin{split}
\textbf{I}_n^2 &= \textbf{I}_n \textbf{I}_n^T\\
&=\left(\sum_{k=1}^m a_{nk} \textbf{u}_k\right)\left(\sum_{k'=1}^m a_{nk'} \textbf{u}^T_{k'}\right)\\
&=\sum_{k=1}^m a_{nk}^2 \lambda_k\;.
\end{split}
\end{equation}

In our discussion of outlier detection below, we will also use the notion of the fractional contribution of eigenimage $k$ to each snapshot $n$ (cf.\ eq.~[\ref{eq:amp}]), which we define as
\begin{equation}\label{eq:aprime}
\begin{split}
a'_{nk} &\equiv \frac{\textbf{I}_n \textbf{u}_k^T}{\sqrt{ \textbf{I}_n^2 \textbf{u}_k^2}}\\
&=\frac{ a_{nk} \sqrt{\lambda_k}}{\sqrt{\sum_{k=1}^m a_{nk}^2 \lambda_k}}\;.
\end{split}
\end{equation}
 
In principle, when we use the above basis set to reconstruct each original image in the ensemble, we need all $m$ eigenimages.
However, depending on the level of fidelity required and on the uniformity  of the images in the set, the PCA decomposition makes it possible for us to reduce the dimensionality of the problem by only using the first few eigenimages to reconstruct an approximation of each of the original images. This approach becomes especially useful when only a few eigenimages are significant and the rest are small. Naturally, the number of eigenimages used to construct the model depends on the particular application and does require judgment. For example, in observational data with real noise, the eigenimage expansion can be terminated when the model begins to overfit the noise. \citet{2010AJ....140..390B} presented a detailed analysis of the optimal way to terminate a PCA expansion given knowledge of the typical S/N ratio of the ensemble images. In characterizing images from simulations that do not include observational noise, the judgment of when to terminate the expansion is one of how much fidelity is required to capture the critical morphology of the image.

Lastly, we emphasize an obvious but important application of PCA. Given a set of eigenimages, the basis can also be used to represent and analyze images that are similar to those in the set used to define the eigenimages but that are not actually in the set itself.  In the present context, this means that a basis constructed from a set of simulated images of an accreting black hole should be able to represent observations of the black hole, if the simulations are sufficiently realistic.

\subsection{Principal Component Analysis in the Fourier Domain}\label{sec:PCA2}

Even though we presented the PCA formalism using a set of images, the data that we ultimately aim to work with are the complex Fourier components of the image, i.e., visibility amplitudes and phases. This is because the EHT is an interferometric array and directly measures the latter quantities. Ideally, we would like to devise a method for characterizing image variability that can be used in both image space and Fourier space and that allows us to move freely between the two. 

From a purely mathematical point of view, the image and Fourier domains are highly symmetric, and it is straight-forward to represent an operation in one domain with a complementary operation in the other domain. In practice, however, the two domains present strongly asymmetric viewpoints. The spatial distribution of radio emission close to the horizon of an accreting black hole is readily formulated and visualized with high-fidelity simulations in the image domain. The observations are obtained in the visibility domain, however, with relatively sparse coverage. Confronting the simulations with the observations requires a sophisticated synthesis of the visibility data into an interpretable form. One path is to use general purpose image reconstruction techniques, but these may suffer from less than optimal use of the expected morphology of the observations. Our approach, instead, will be to develop a basis directly in the visibility domain that encodes the expected behavior of the source as informed by simulations. We thus need to understand how the PCA basis of the simulations will relate to their visibilities. In this section, we show that the visibilities of the principal components of the simulation images are in fact the same as the principal components of the visibilities of these images. 

We define the 2D discrete Fourier transform of an image as
\begin{equation}
\tilde{I}_{\alpha} = \sum _{i=1}^{N^2} F_{\alpha i} I_{i}\;,\qquad \alpha=1,...,N^2\;.
\end{equation}
Here, in order to account for the folding of the images into one-dimensional vectors, we have written the discrete Fourier operator in the compact form
\begin{equation}
F_{\alpha i} = e^{\frac{-2\pi \mathrm{i}}{N}[j\beta+k\delta]}\;,
\label{eq:Fourier}
\end{equation}
where the indices $j, k, \beta$, and $\delta$ in the right hand side of this relation can be evaluated from the indices $\alpha$ and $i$ via the relations
\begin{eqnarray}
k&=&\left[\left(i-1\right) \mod N\right]+1\;,\\ 
j&=&\frac{i-k}{N}+1
\end{eqnarray}
and
\begin{eqnarray}
\delta&=&\left[\left(\alpha-1\right) \mod N\right]+1\;,\\ 
\beta&=&\frac{\alpha-\delta}{N}+1\;.
\end{eqnarray}
Note that in equation~(\ref{eq:Fourier}) we used the symbol $\mathrm{i}$ for the imaginary number to distinguish it from index $i$.

The Fourier transform of matrix $A$ is simply
\begin{equation}
\tilde{A}_{\alpha n} = \sum _{i=1}^{N^2} F_{\alpha i} A_{in}
\end{equation}
and we define the $m \times m$ matrix $ L^\prime \equiv\tilde{A}^T\tilde{A}$ as 
\begin{equation}
L^\prime\equiv\tilde{A}^T\tilde{A}
=
\begin{bmatrix}
    \tilde{I}_1 \tilde{I}_1 & \tilde{I}_1 \tilde{I}_2  & \cdots  & \tilde{I}_1 \tilde{I}_m \\
    \tilde{I}_2 \tilde{I}_1 & \tilde{I}_2 \tilde{I}_2  & \cdots  & \tilde{I}_2 \tilde{I}_m \\
    \vdots & \vdots & \ddots & \vdots \\
    \tilde{I}_m \tilde{I}_1 & \tilde{I}_m \tilde{I}_2  & \cdots  & \tilde{I}_m \tilde{I}_m \\
\end{bmatrix}.
\end{equation}
Our goal here is to show that this matrix is the same as $L$, i.e., that $L^\prime=L$.

We write the vector product that appears in each element of matrix $L^\prime$ as 
\begin{equation}
\begin{split}
\tilde{\mathbf{I}}_f \tilde{\mathbf{I}}_g &= \sum _{\alpha=1}^{N^2} \tilde{I}_{\alpha f} \tilde{I}^*_{\alpha g} \\
&= \sum _{\alpha=1}^{N^2} \left(\sum_{i=1}^{N^2} F_{\alpha i} I_{if}\right) \left(\sum_{i'=1}^{N^2} F^*_{\alpha i^\prime} I_{i^\prime g}\right).
\end{split}
\end{equation}
Because $I_{if}$ and $I_{i'g}$ do not depend on $\alpha$, we can rearrange the above equation as follows
\begin{equation}
\begin{split}
\tilde{\mathbf{I}}_f \tilde{\mathbf{I}}_g&= \sum_{i=1}^{N^2}  I_{if}\sum_{i'=1}^{N^2} I_{i'g}\left(\sum _{\alpha=1}^{N^2} F_{\alpha i}F^*_{\alpha i'}\right).
\end{split}
\end{equation}
The term in parenthesis above is the 2D Fourier transform of a constant and is equal to $\delta(i-i')$ such that
\begin{equation}
\begin{split}
\tilde{\mathbf{I}}_f \tilde{\mathbf{I}}_g&= \sum_{i=1}^{N^2} I_{if}\sum_{i'=1}^{N^2}  I_{i'g}
\delta(i-i')\\
&= \sum_{i=1}^{N^2} I_{if}I_{ig}=\mathbf{I}_f \mathbf{I}_g\;.
\end{split}
\end{equation}
Therefore, each element of $L$ is equal to that of $L'$ so their eigenvectors and eigenvalues must also be equal ($L'\textbf{v}_{\gamma}=\lambda_{\gamma}\textbf{v}_{\gamma}$).
\begin{figure*}[t!]
\centering
\includegraphics{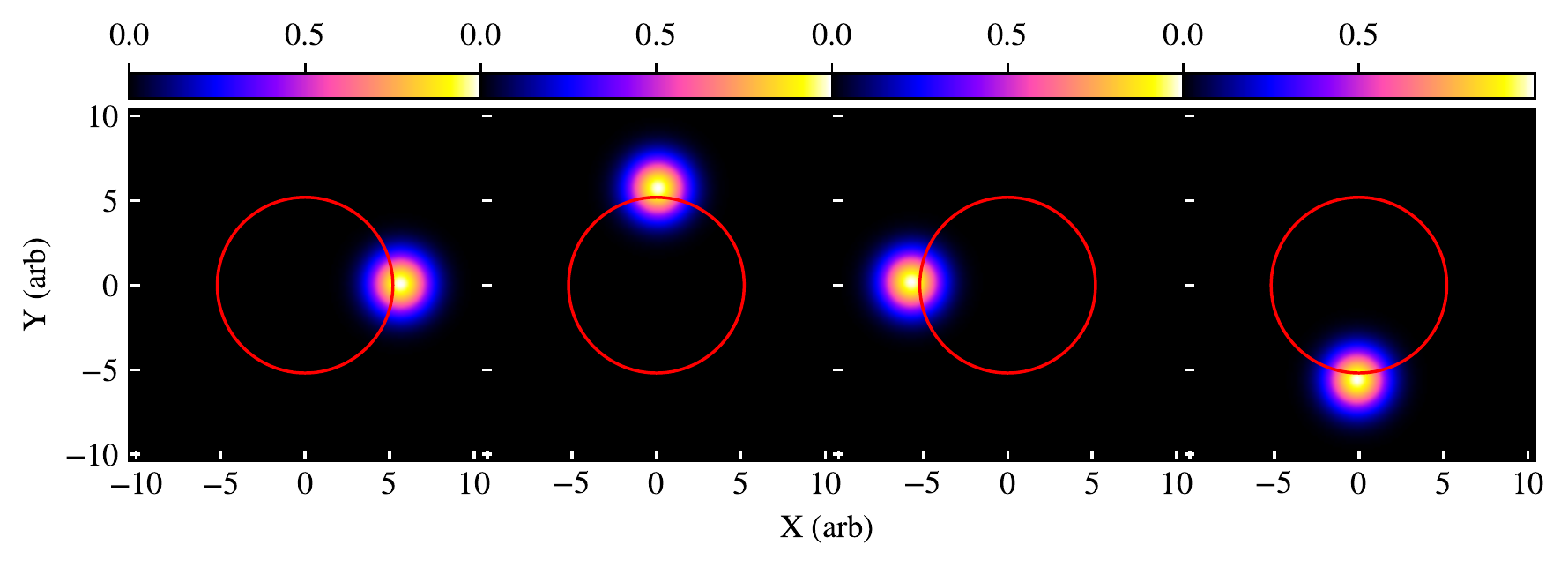}
\caption{Example snapshots from a simple model of a Gaussian spot moving on a circular path. Here the red circle indicates the approximate trajectory of the center of the Gaussian spot. The linear scale of the image is arbitrary. We present PCA analysis of realistic GRMHD simulations later in the paper, but
this simple example is useful for understanding how PCA decomposition of the simulations work.}
\label{fig:circ_snaps}
\end{figure*}

\begin{figure*}[t!]
\centering
\includegraphics{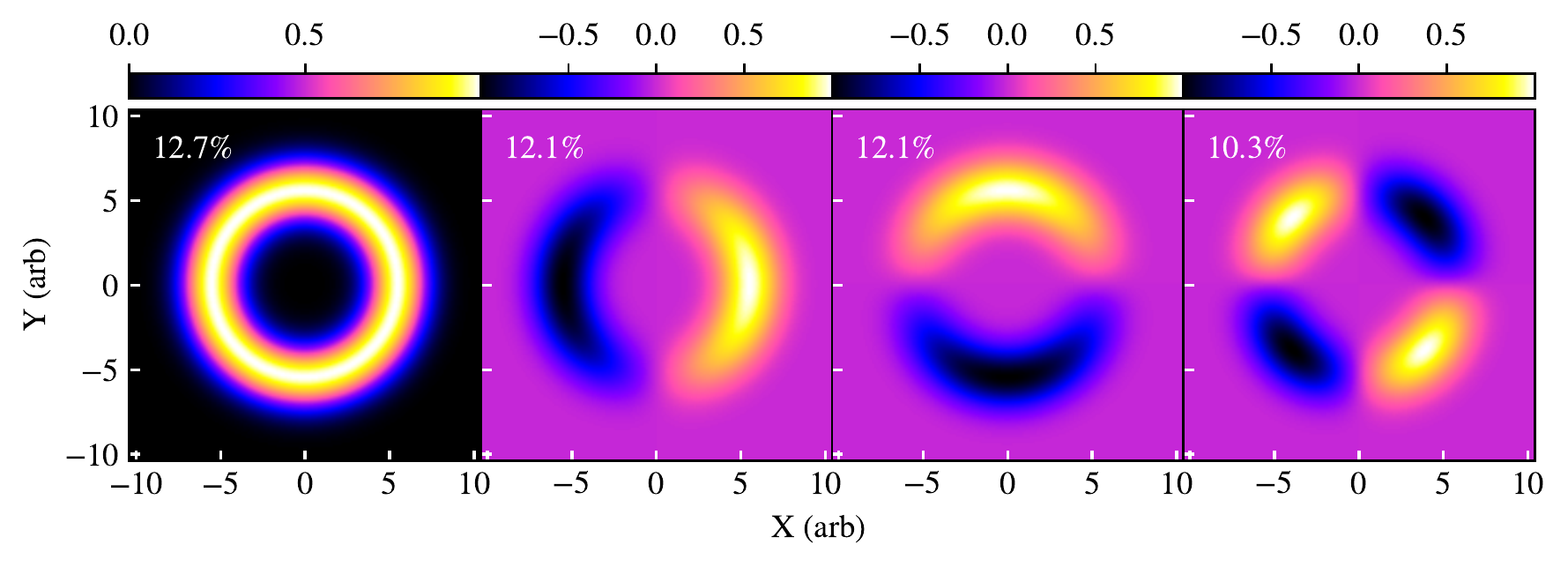}
\caption{The first 4 components of the PCA decomposition of the Gaussian spot moving on the circular path shown in Figure \ref{fig:circ_snaps}. The eigenvalues which correspond to these four components are shown in the top left of each panel, respectively. Note that in this figure, and in all figures of principal components in the rest of the paper, each component has been normalized independently so fluxes cannot be compared between different components.}
\label{fig:comp_circ}
\end{figure*}

We now define the covariance matrix for the visibilities in analogy to that of the images as 
\begin{equation}
C' = \tilde{A}\tilde{A}^{*T}.
\end{equation}
As before, we can find the eigenvectors of $C'$ by diagonalizing $L'$ [see equation (\ref{eq:AATA_CA})], which demonstrates equivalently that $\tilde{A} \textbf{v}_{\gamma}$ are eigenvectors of $C'$ with eigenvalues $\lambda_{\gamma}$. In order to complete our proof, we must show that the eigenvectors $\tilde{A} \textbf{v}_{\gamma}$ are equal to the Fourier transform of the eigenvectors $A \textbf{v}_{\gamma}$. In other words, we must show that the principal components of the set of visibility maps is equal to the visibilities of the principal components of the set of images. This can be seen by evaluating each component of the eigenvector $\tilde{A} \textbf{v}_{\gamma}$ as
\begin{equation} \label{eq1}
\begin{split}
(\tilde{A} \textbf{v}_\gamma)_\alpha &=\sum_{l=1}^m\tilde{A}_{\alpha l} (\textbf{v}_{\gamma})_l =\sum_{l=1}^m \left(\sum_{i=1}^{N^2} F_{\alpha i} A_{il}\right)(\textbf{v}_{\gamma})_l\\
&= \sum_{i=1}^{N^2} F_{\alpha i} \left(\sum_{l=1}^m A_{il}(\textbf{v}_{\gamma})_l\right)=\tilde{(A \textbf{v}_\gamma)}_\alpha \;.
\end{split}
\end{equation}
Therefore, the visibilities of the principal components are indeed equal to the principal components of the visibilities. We have thus shown that the PCA basis can be developed in one domain, such as image space, but is readily applied in the complementary Fourier (or visibility) domain. This will allow us to use PCA to compare and possibly fit EHT data to simulations directly in the visibility domain.  The maximally compact basis provided by the PCA approach may be well-suited to address the sparse coverage of the visibilities.

\section{An Example of Principal Component Analysis}\label{sec:PCA3}
\begin{figure}[t!]
\centering
\includegraphics{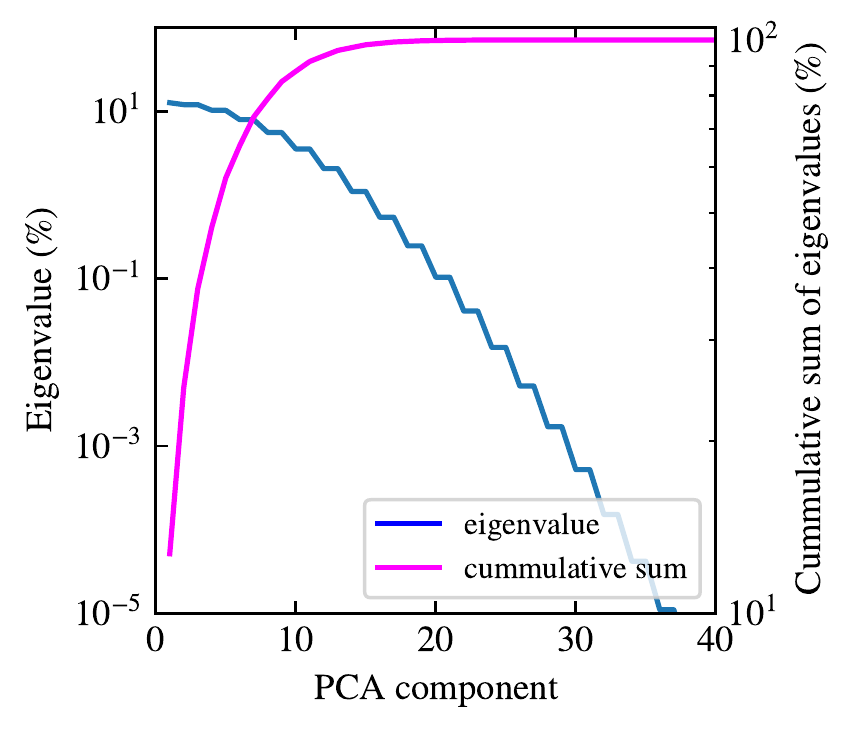}
\caption{{\em (Blue curve)\/} The spectrum of eigenvalues for the PCA decomposition of the Gaussian spot shown in Figure \ref{fig:circ_snaps}; the eigenvalues have been normalized such that they sum up to 100\%. The step-like features in this spectrum are present because the high degree of symmetry in this model causes the principal components to come in pairs with very similar eigenvalues (see the second and third panels of Fig.~\ref{fig:comp_circ}). {\em (Magenta curve)} The cumulative sum of the eigenvalues. Note that only the first $\sim 40$ of the 1080 components are shown and that the first 10 components contain 88\% of the structural information.}
\label{fig:spec_circ}
\end{figure}

To elucidate how PCA works in practice, we present, in this section, a simple example that is easy to calculate and understand. We consider a Gaussian spot moving along a circular path and simulate 1080 snapshot images, as the spot completes an integer number (3) of orbits. Figure \ref{fig:circ_snaps} shows a few example snapshots from this model. We calculate the principal components of this image set using the PCA formalism and show in  Figure~\ref{fig:comp_circ} the first few principal components. We also show in  Figure~\ref{fig:spec_circ} the spectrum of eigenvalues we obtain for this model. 

The first principal component, which has the largest flux variance ($\sim 12.7\%$), amounts to the average image of the various snapshots (modulo a normalization constant), i.e., it represents a ring surrounding the circular path with a width comparable to the width of the Gaussian spot. This is not a general property of a PCA decomposition but is exact in the particular example discussed here and approximately correct in the PCA decomposition of the black-hole images we will discuss in the next section. In the example of the orbiting Gaussian spot, all terms in each row of the $m\times m$ matrix $L$ also appear in each other row of the same matrix, but displaced at different columns. This is true because the product of any two images in the ensemble depends only on the relative positions of the Gaussian spots in the two images. In other words, the sum of the elements of each row of matrix $L$, i.e., $\sum_{n=1}^{m}L_{in}$ is constant. One of the eigenvectors of a matrix with elements that obey this property is a vector that has all elements equal to one (or actually any constant, depending on how the eigenvector is normalized), i.e., $\textbf{v}_1=[1\; 1\; 1\; 1\; ...\; 1]$. The eigenvector of the covariance matrix $C$ that corresponds to this eigenvector of the matrix $L$ is then (see eq.[\ref{eq:eigenimages}])
\begin{equation}
\textbf{u}_1=A \textbf{v}_1=\sum_{n=1}^m \textbf{I}_m\;,
\end{equation}
which (modulo a normalization constant) is nothing but the average image of the ensemble.

\begin{figure*}[t!]
\centering
\includegraphics{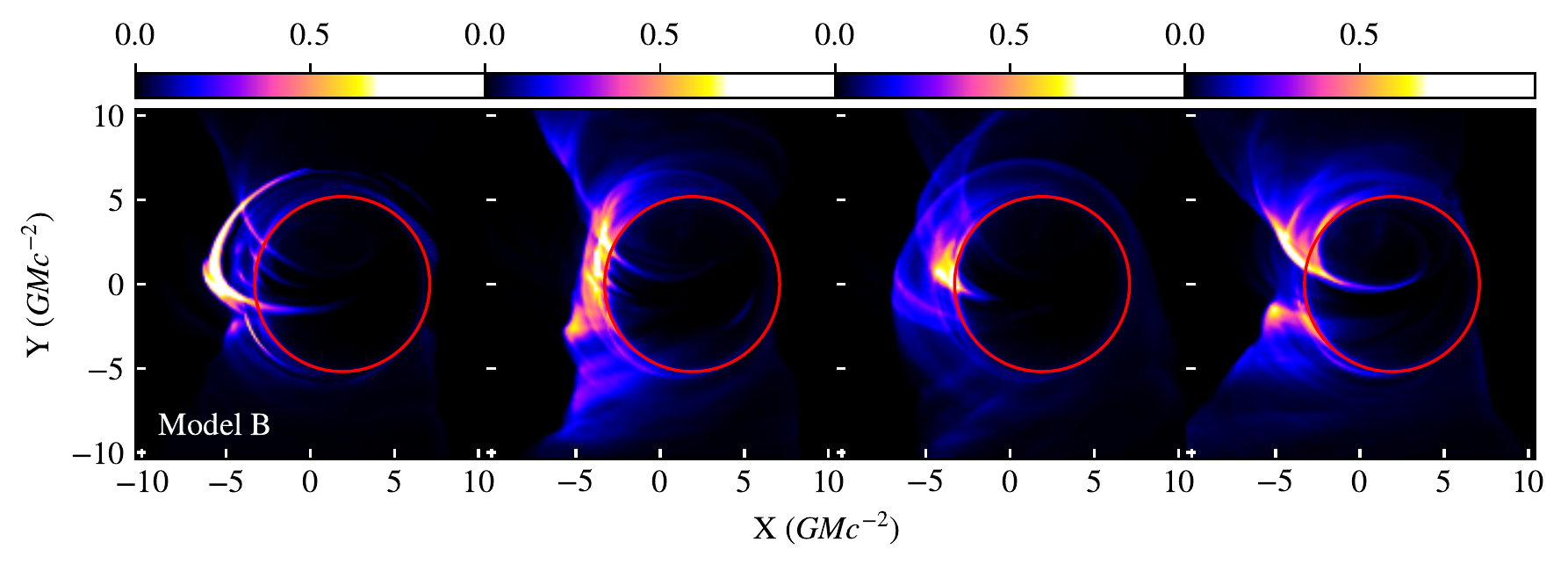}
\caption{Four example snapshots of the ensemble of black-hole images computed using Model~B at a wavelength of 1.3~mm (see text for description of the model and of the simulations). None of these snapshots correspond to an instant with a significant flare in the flux of the black hole.  For this and the following figures, the peak flux in each panel has been normalized to unity so changes in overall flux have not been preserved. The original images span 32 $GM/c^2$ on each side and the full size images are used in all calculations; however, we choose to show only the innermost $\sim 20 GM/c^2$ in the figures throughout the paper so that the black hole shadows are easy to distinguish. The red circles in the figures correspond to the expected size and location of the black hole shadow for each particular model. The location of the circle relative to the center of the image depends on the black hole spin and is not necessarily centered on the location of the black hole itself. }
\label{fig:snaps}
\end{figure*}

Most eigenimages, other than the lowest-order one, have pixels with significantly positive and negative brightness, as one would expect from an eigenvector decomposition. However, because all images are positive definite and the lowest-order eigenimage is the average of the ensemble of images, it follows that the lowest-order eigenimage is also positive definite. Moreover, because of the symmetry of the ensemble of images, components 2 and 3 of the spectrum of eigenimages (in Figure \ref{fig:comp_circ}) differ only by a rotation. The eigenvalue connected to each eigenimage is related to its variance (see equation~[\ref{eq:eigenv}]) and, hence, these two components correspond to the same eigenvalue. This behavior persists with higher components such that components come in pairs with similar eigenvalues. This creates the step pattern present in Figure \ref{fig:spec_circ}. 

Here and in all simulations of black-hole images discussed in the sections below, the typical values of the amplitudes $a_{nk}$ are very similar between different eigenimages, i.e., the typical values and distribution of the amplitudes $a_{nk}$ depend weakly on $k$. Because of this and the fact that the eigenimages are normalized according to their eigenvalues (see eq.~[\ref{eq:eigenv}]), the spectrum of eigenvalues, such as the one shown in Figure~\ref{fig:spec_circ}, matches very closely the spectrum of the relative contributions of each eigenimage to the reconstruction of any of the $m$ images in the ensemble. As a result, we can use the spectrum of eigenvalues as a proxy for investigating the relative contribution of each eigenimage to the reconstruction of a typical image in the ensemble (see below for outlier detection). This is the reason why we normalize all eigenvalues so that they sum to unity (see eq.[\ref{eq:suml}]) and we often quote them as percentages.

In the example of the circulating Gaussian spot that we discuss in this section, the eigenvalues drop dramatically after the first few, indicating that only a few components would be sufficient to reconstruct the original images. Specifically, under the assumption that the parameters $a_{nk}$ are independent of $k$, we conclude that the first 10 (out of 1080) components account for $\approx 88\%$ of the structures present in the ensemble of images, while $\sim 25$ components account for nearly all of it. For this particular example, it is straightforward to understand why it takes only a small fraction of the eigenimages in order to reconstruct most of the structures seen in each of the images in the ensemble by estimating the number of substantially different images that are present in the ensemble. For the parameters used in this model, the FWHM of the Gaussian spot subtends $\sim 13^{\circ}$ as viewed from the center of the circular path. Therefore, the full circular trajectory can be decomposed into 28 distinct Gaussian spots that are (mostly) not overlapping. In other words, there are only 28 ``resolution'' elements in the circular trajectory and, therefore, the contribution of all but the first $\sim 28$ eigenimages can only be negligible.

\section{Principal components analysis of simulated black hole images}\label{sec:PCAsims}

We now apply PCA to a set of simulated black hole images at the 1.3~mm wavelength of observations for the EHT.  We focus on three of the five best-fit models \citep{chan2015, 2017ApJ...844...35M,  Medeiros2018} that were calibrated to reproduce the broadband spectrum of \sgra and the size of the 1.3~mm emission region inferred by early EHT observations~\citep{2008Natur.455...78D}. These three models span the range of image morphologies and structural variability that we encountered in all our simulations. Specifically, Model B has a 1.3~mm image that is dominated by the accretion disk region and resembles a crescent shape, Model C has a 1.3~mm image that is dominated by the base of the jet funnel, while Model D has a 1.3~mm image that is a combination of both the base of the funnel and the disk (here we use the nomenclature of \citealt{chan2015} to label the models). 

The simulations were generated by performing time-dependent general relativistic magnetohydrodynamic (GRMHD) simulations using the 3D \texttt{HARM} code \citep{2003ApJ...589..444G, sadowski_fix, 2013MNRAS.436.3856S} and by carrying out radiative transfer and geodesic ray tracing calculations on the simulation outputs using \texttt{GRay} (\citealt{2013ApJ...777...13C}). Because \texttt{GRay} is a massively parallel GPU based code, we were able to generate images as a function of time with high spatial and temporal resolution, for a large number of simulations, while varying the black hole spin, the geometry of the magnetic field, and the plasma model (see \citealt{2015ApJ...799....1C} for a detailed description).

\begin{figure*}[t!]
\centering
\includegraphics{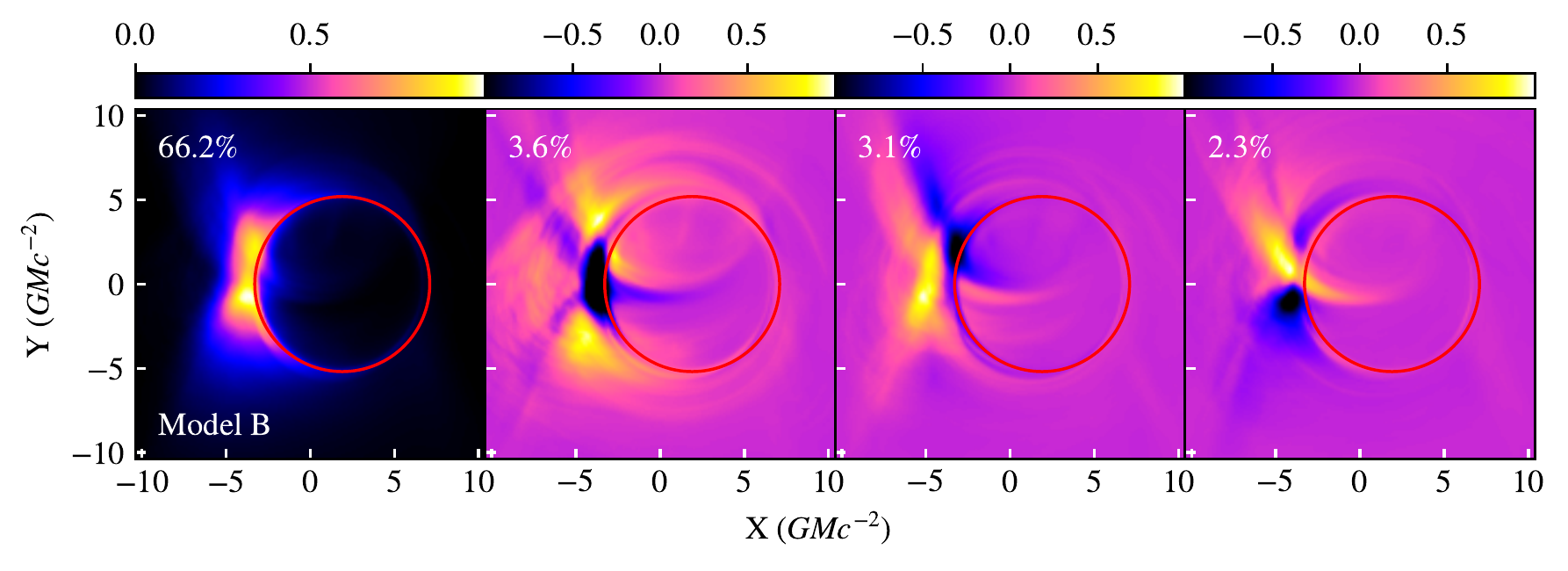}
\includegraphics{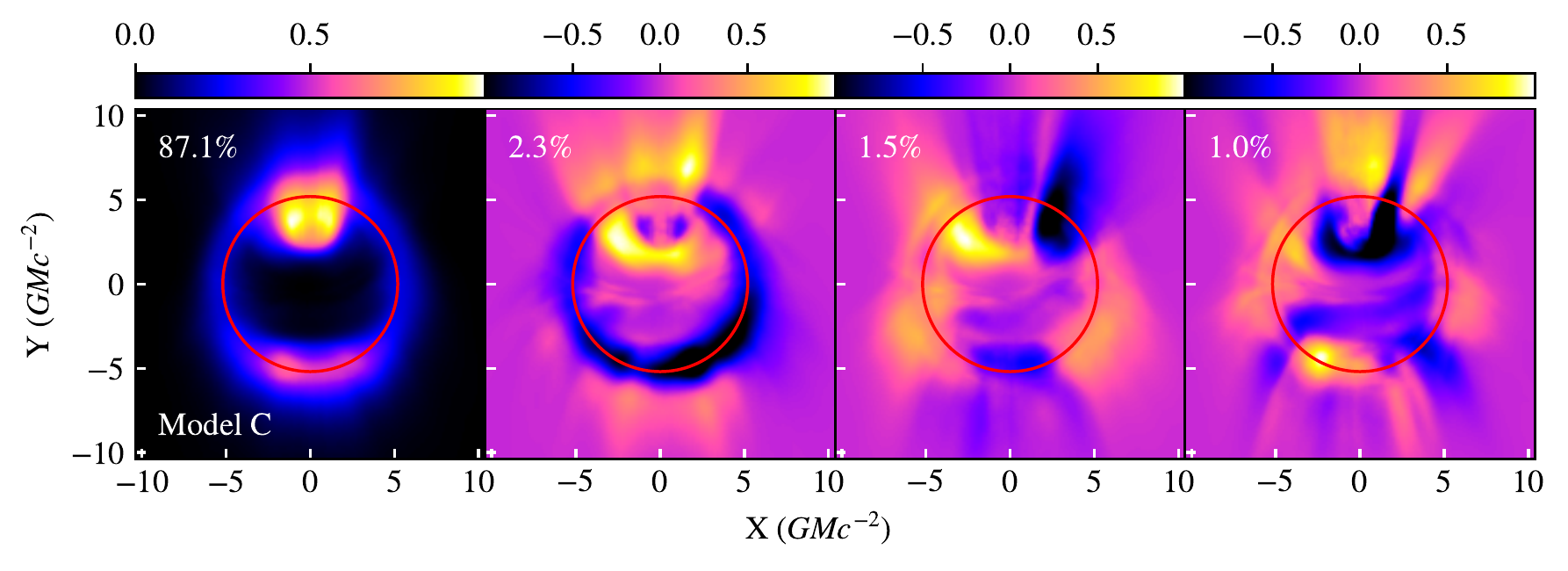}
\includegraphics{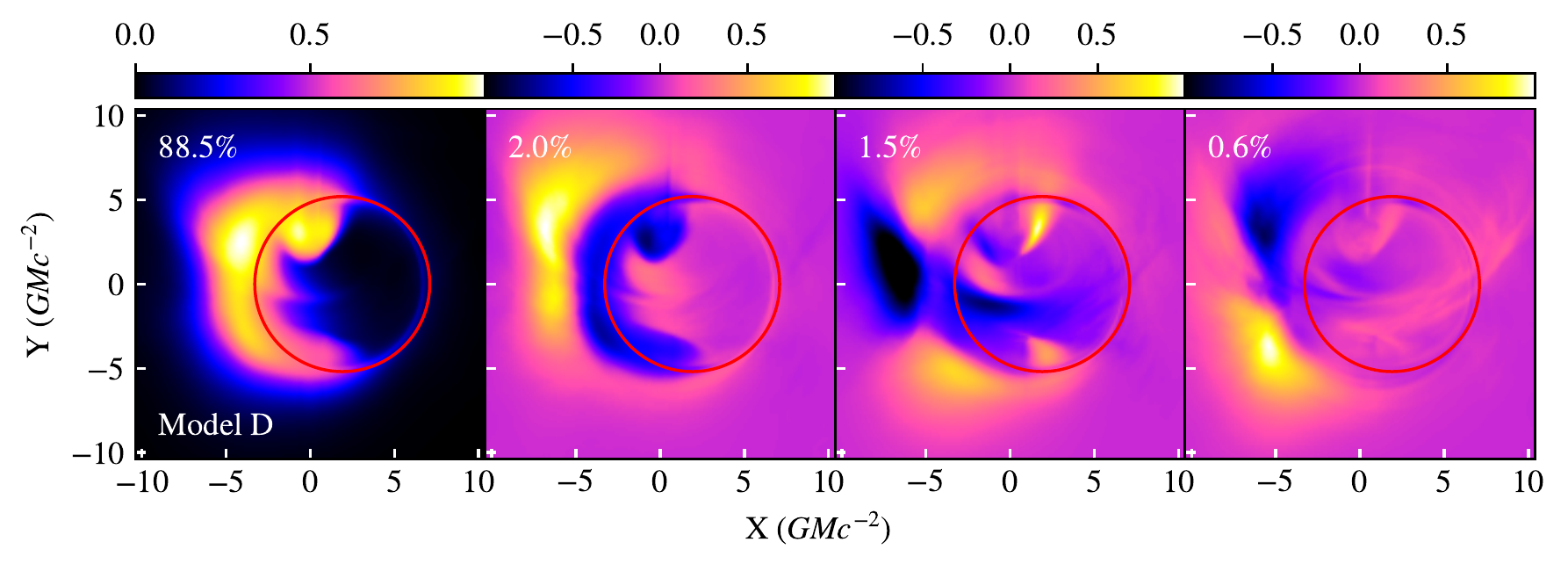}
\caption{The first four principal components and their corresponding eigenvalues for the three GRMHD simulations (models B, C, and D) described in the text. The principal component for each simulation (leftmost panel) is approximately equal to the average image from the simulation (see Figure 1 in \citealt{2017ApJ...844...35M}). 
}
\label{fig:comp}
\end{figure*}

We show in Figure~\ref{fig:snaps} four example snapshots from Model B highlighting the structural variation in the emission region that is prominent in this model. Hereafter, when displaying images of black holes, we will measure all lengths in units of the gravitational radius $GM/c^2$, where $M$ is the mass of the black hole and $G$ and $c$ are the gravitational constant and speed of light, respectively. The radius of the black-hole shadow is approximately equal to 5 gravitational radii while the center of the shadow is displaced with respect to the center of gravity, depending on the spin of the black hole (see, e.g., \citealt{2013ApJ...777...13C}).

We perform PCA on the three simulations described above following the procedure outlined in \S 2. Each image set consists of 1024 images corresponding to the number of snapshots obtained from the accretion flow simulations that span $\approx 60$ hours. In Figure \ref{fig:comp}, we show the first 4 eigenimages and their respective eigenvalues for the PCA decomposition of the three models. For all models, the first eigenimage (left) is similar to the time average of the ensemble of images (see Figure 1 in \citealt{2017ApJ...844...35M} for a comparison with the time averaged images of these simulations). This is true because all images have a dominant structure (i.e., a crescent or the footprints of the funnel), on top of which the variability of the accretion flow introduces sub-dominant perturbations. As a result, the correlations between the various snapshots are very similar to each other and the arguments given in \S3 for the dominant eigenimage apply here as well, but only approximately.

Although PCA is a purely mathematical tool and is agnostic about the physics of the system, some of the components do appear to have identified important physical features. For example, the second component in Model~B shown in the top row of Figure~\ref{fig:comp} appears to have identified a region of the Doppler boosted accretion disk (center of the crescent shape) that is very close to the black hole shadow and is highly variable in the simulations (see e.g., the second and third panels of Figure \ref{fig:snaps}). The third and fourth components of Model B appear to be tracing the Doppler boosted walls of the funnel region. This also matches the behavior that can be seen directly in the simulation, where the relative brightness of the wall of the funnel region is highly variable (see, e.g., the fourth panel of Figure \ref{fig:snaps}). Note that, due to the 60 degree inclination of the observer relative to the spin axis of the black hole in these simulations, the base of the funnel appears to come from within the black hole shadow but is actually positioned between the observer and the black hole.

\begin{figure}[t!]
\centering
\includegraphics{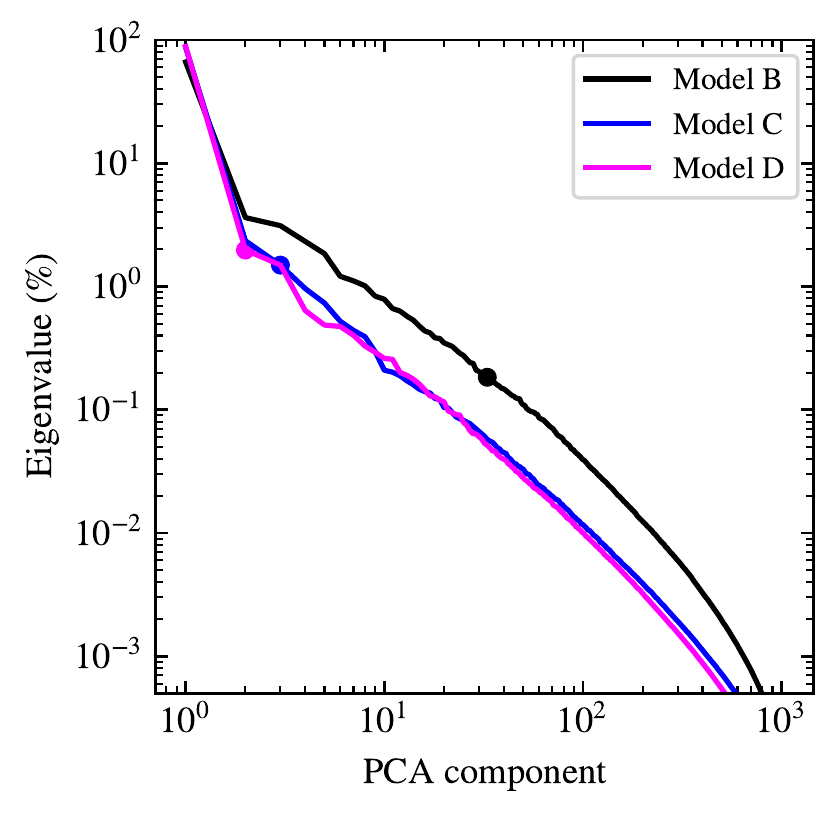}
\caption{The eigenvalue spectra of the PCA decomposition of the images from the three GRMHD simulations we consider. The filled circles along each spectrum indicate the number of PCA components that are required to account for $90\%$ of the image structures. The rapid decay of the eigenvalue spectrum indicates that PCA can be used to reduce significantly the dimensionality of the ensemble of images that arise in these simulations.}
\label{fig:spec_mod}
\end{figure}

In Model C (middle row), the second component highlights the edge of the black hole shadow while the other two components show various ways in which the structure of the emission at the base of the funnel varies. The PCA decomposition for Model D (bottom row) shows some features of Model B, i.e., that a crescent shape is present, and also some features of Model C, i.e., that the base of the funnel is an important variable feature in the image.

In Figure \ref{fig:spec_mod}, we plot the spectra of eigenvalues of the PCA decomposition of the images from the three models. Unlike the example of the Gaussian spot discussed in \S \ref{sec:PCA3}, the eigenvalue of the first principal component in all three models overwhelms that of the remaining components. For example, the eigenvalue of the first component of Model D corresponds to $\sim 89\%$, whereas the eigenvalue of the second component drops to only $\sim 2\%$. In other words, under the assumptions discussed in \S3, only the first two components (out of 1024) are required to account for $90\%$ of the structures in the images from Model D and only the first three components are required to reach the same level for Model C. This result indicates that PCA can be extremely useful in reducing the dimensionality of the images that arise in these GRMHD simulations and that only the first few components are needed to preserve the majority of the image structure. 


Model B differs somewhat from the other two models in this regard. The eigenvalue that corresponds to the first component is equal to $\sim 66\%$, i.e., it is $\sim 20\%$ less than in the other two models. Correspondingly, as many as 33 components are required to account for $90\%$ of the structure seen in the images for model B, showing that this model contains significantly more structural variability than the other models. This is in agreement with the findings reported in \citet{chan2015} and  \citet{2017ApJ...844...35M, Medeiros2018}, where the higher level of flux variability and flaring behavior was attributed to structural changes rather than to simple brightness fluctuations. Nevertheless, even for such a simulation that shows more significant structural variability, the required number of components (33) is significantly smaller than the total number of images, making PCA useful for dimensionality reduction. We explore this result further in the following section.

It is intriguing that despite the differences in the relative importance of the first $\sim 10$ eigenvalues, the eigenvalue spectrum declines with the same slope for the higher components in all models. This suggests a common origin for the slope of the eigenvalue spectrum, which we will explore in detail in \S5. 

\begin{figure*}[t!]
\centering
\includegraphics{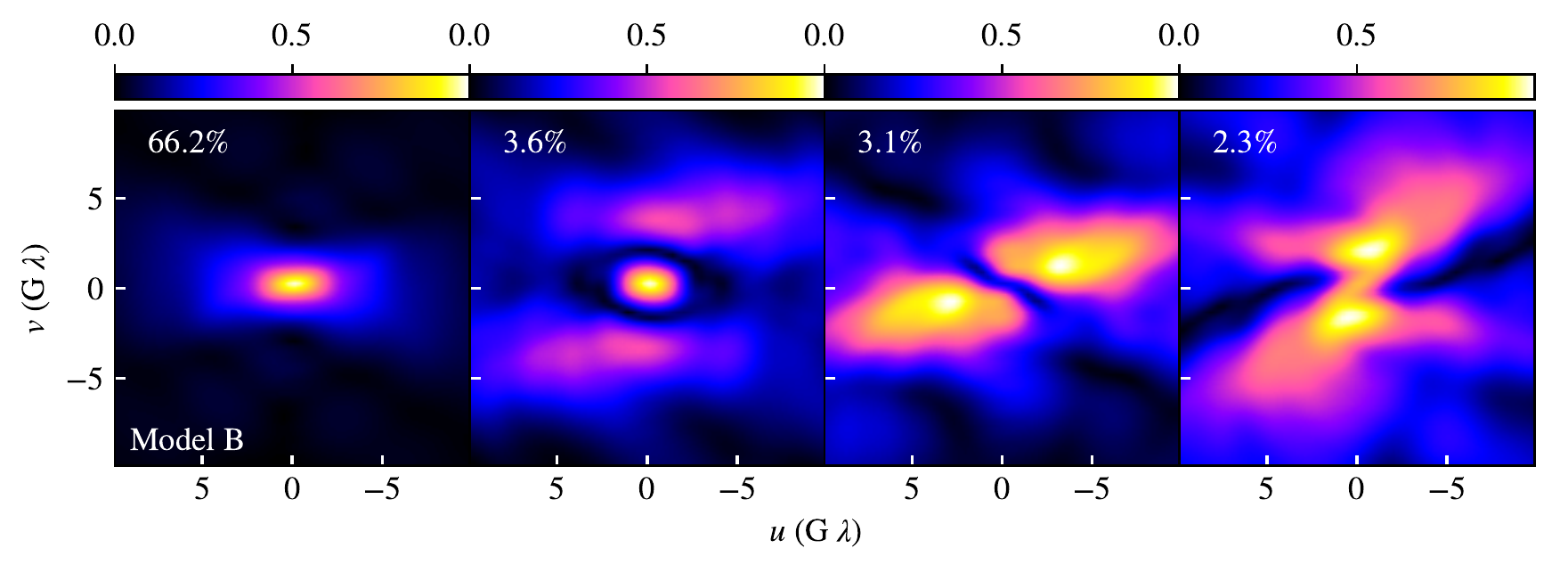}
\includegraphics{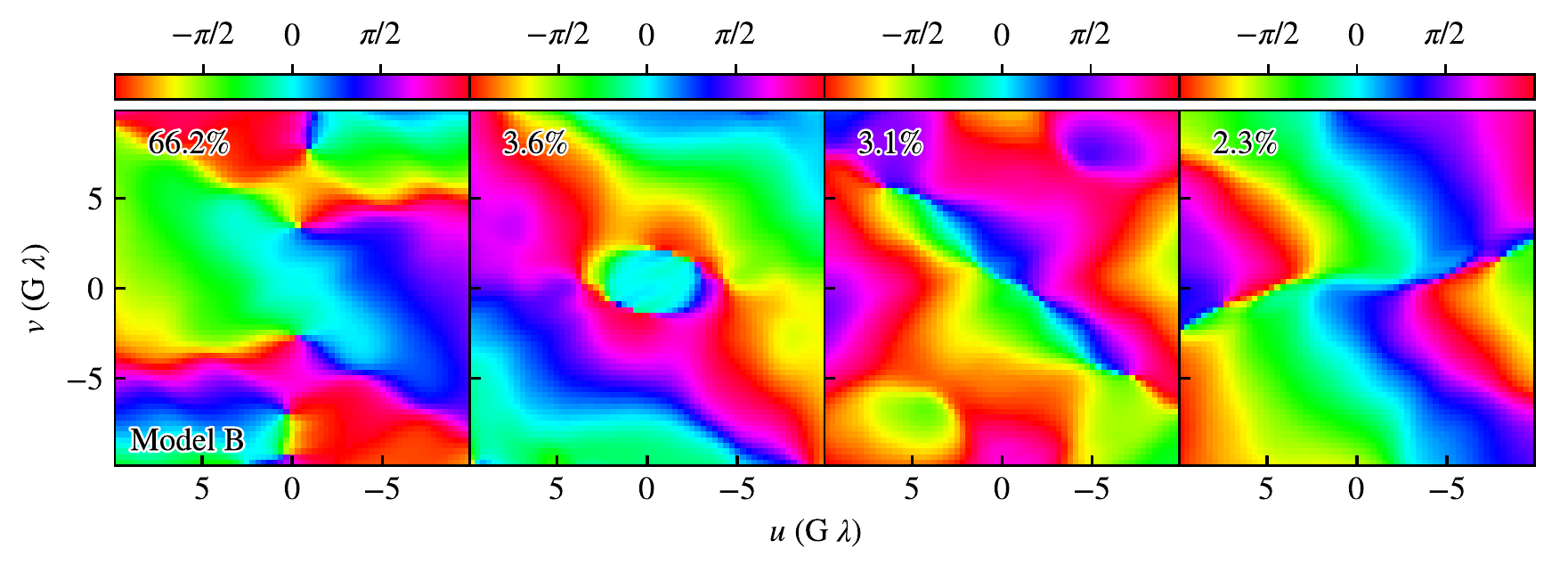}
\caption{Visibility amplitudes (top row) and visibility phases (bottom row) of the first four components of the PCA decomposition of Model B (cf.\ the top row of Figure \ref{fig:comp}). Higher components contribute significantly at increasingly longer baselines.}
\label{fig:VA_VP_B}
\end{figure*}

Finally, we also apply the PCA analysis directly on the complex visibilities of our image set, which are the components of the 2D Fourier transform of each image. As we showed in \S2.2, we can either calculate the complex visibility maps for each image in our ensemble and then perform PCA or  calculate directly the complex visibility maps for each PCA component of the ensemble of images; the results will be identical. Given that the images correspond to vectors with real elements whereas the visibilities correspond to vectors with complex elements, we follow the second procedure, which is easier to implement. In Figure \ref{fig:VA_VP_B}, we show the first four PCA components of the visibility amplitudes and visibility phases of Model B. As expected, the structures of the visibility amplitudes and phases changes significantly between these four components. In fact, the visibility amplitudes of higher components have more power at longer baselines, which is a direct consequence of the fact that they contain smaller scale structures.

\section{Dimensionality reduction and outlier identification}\label{sec:reconstruction}

As we discussed in the previous section, only the first few PCA components are required to account for the majority of the structure seen in the images from each GRMHD simulation. Components with smaller eigenvalues both contribute less to the brightness of each pixel in the image (see discussion at the end of \S3) and account primarily for small-scale structures (see Figure~\ref{fig:VA_VP_B}).  
As alluded to in the introduction, this conclusion (often referred to as dimensionality reduction) also offers the possibility of using a small number of measurements, such as those possible with the sparse coverage of the EHT array, to reconstruct the persistent image of a black hole and, therefore, extract the information that is most relevant for detecting its shadow. To explore the idea of using the first few components to describe the persistent structure from the variable flow, we calculate and compare reconstructions of images from the simulations using the first few components of the PCA decomposition to the original images.

\begin{figure*}[t!]
\centering
\includegraphics{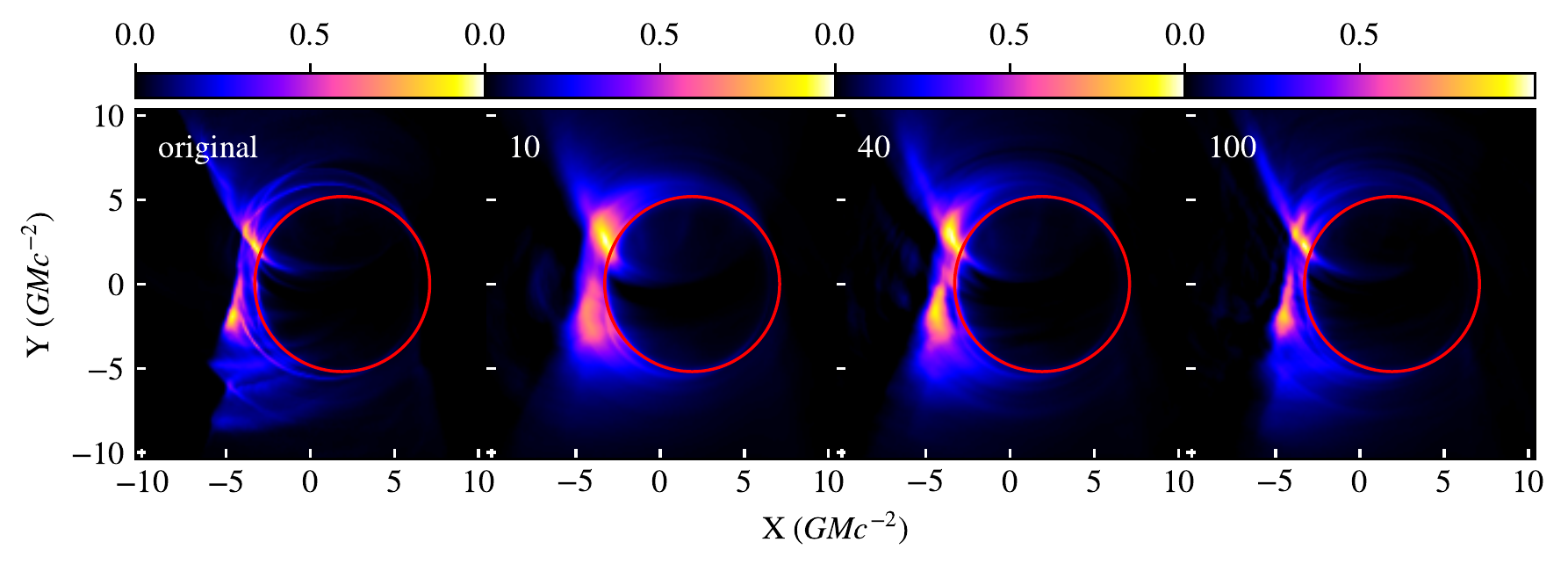}
\caption{The left panel shows a typical snapshot from Model B. The three right panels show the same snapshot from Model B but reconstructed using only the first 10, 40, and 100 components from the PCA decomposition. The reconstruction using only the first 10 components smooths over the fine scale structure but faithfully reproduces the overall brightness distribution of the full image.
}
\label{fig:recon}
\end{figure*}

\begin{figure*}[t!]
\centering
\includegraphics{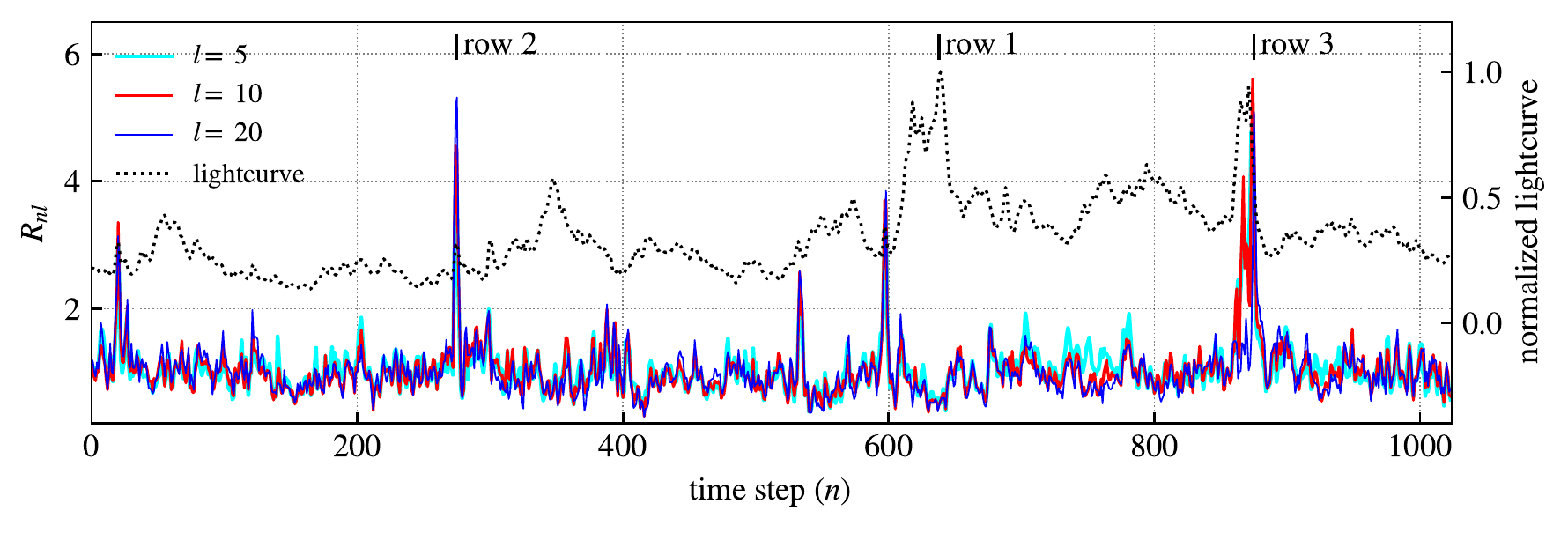}
\caption{The dotted black line shows the normalized light curve for Model B. The cyan, red, and blue curves correspond to the quantity $R_{nl}$ for $l=5$, 10, and 20, respectively. The peaks in the cyan, red, and blue curves indicate time instances that cannot be adequately reconstructed by only the first few components and are, hence, identified as outliers. The three time instances identified as `row~1',`row~2', and `row~3' correspond to the images that are shown in Figure~\ref{fig:outliers2}.}
\label{fig:outliers1}
\end{figure*}

Figure~\ref{fig:recon} shows an example snapshot from the Model B simulation compared to its reconstruction using the first 10, 40, and 100 out of the 1024 PCA components. Although the reconstructions with only a small number of components do not reproduce the finer details of the images, they do capture their overall structure. The fidelity of reconstruction naturally increases as more components are added. The number of components we may choose to keep in a particular reconstruction and, hence, the degree of dimensionality reduction will naturally depend on the goal of the reconstruction. Nevertheless, even at a qualitative level, this figure suggests that dimensionality reduction by factors of 10 to 100 may be achievable in characterizing black-hole images with PCA.

The snapshot in Figure~\ref{fig:recon} is typical and, therefore, can easily be reconstructed using only the first few PCA components. However, there may be snapshots within a given simulation that are much harder to reconstruct because the structure of the image is unusual compared to the rest of the ensemble. For the purposes of this work, we will define an outlier as an image that cannot be easily reconstructed by the first few (or a ``typical" number of) eigenimages. As we will show below, PCA allows us to devise an algorithmic approach for outlier detection. When we apply PCA to numerical simulations, this outlier detection will enable us to efficiently identify instances where rare and episodic events occurred in the simulation, e.g., a flare in the emission properties of the accretion flow. When we apply PCA to observational data, detecting outliers will allow us to identify similar episodic events that may be caused by physical phenomena or data corruption.

\begin{figure*}[t!]
\centering
\includegraphics{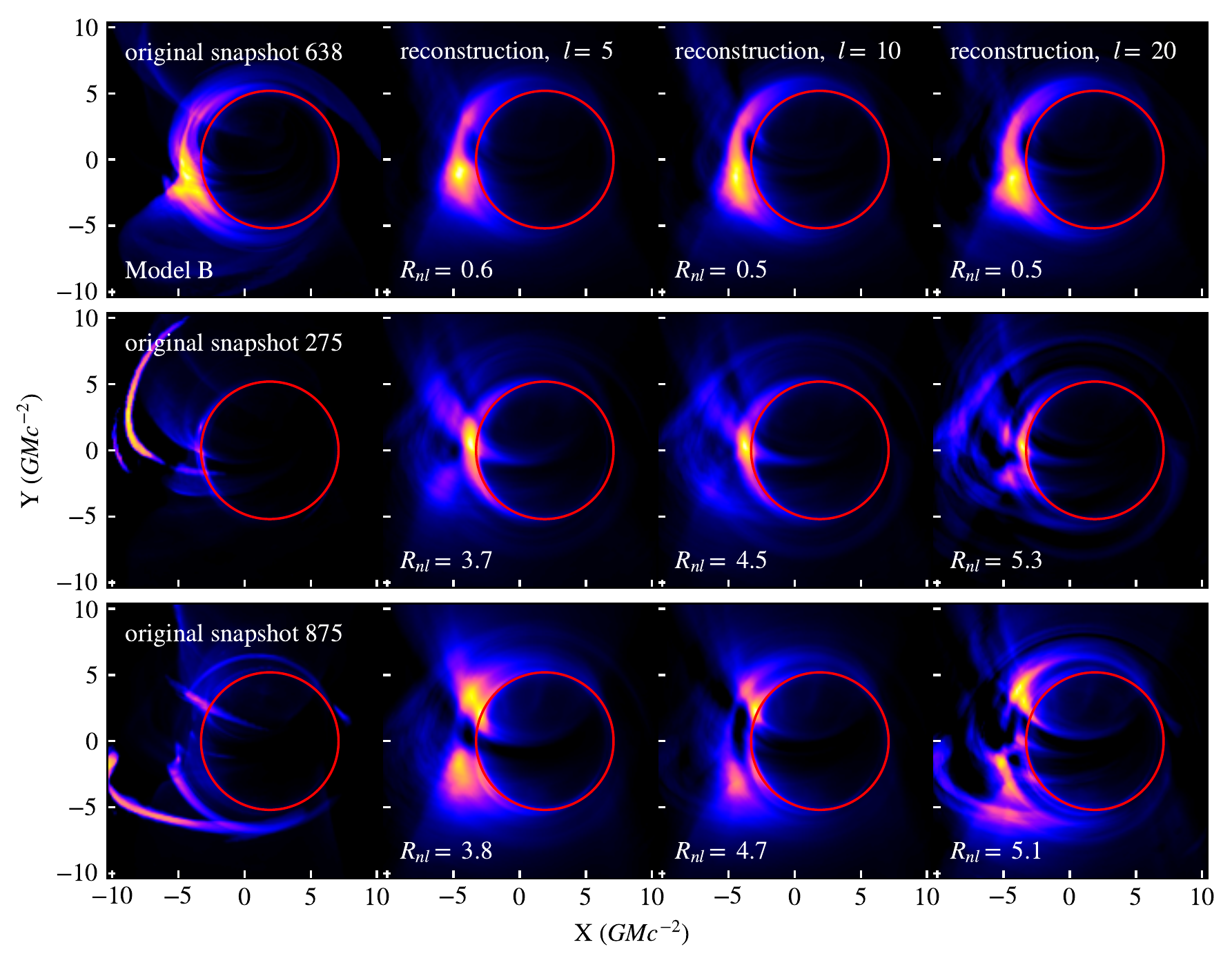}
\caption{The leftmost panels show three snapshots from the ensemble of images calculated for Model~B. In each row, the three consecutive panels show reconstructions using the first 5, 10, and 20 principal components. The top row corresponds to a time instance that is not identified as an outlier, but corresponds to a large flux excursion from the accretion flow (this time step is denoted by ``row 1" in Figure \ref{fig:outliers1}). Note that this image is well-fitted with the first 20 eigenimages, and thus has a small $R_{nl}$ value. The second row corresponds to a time step which is identified as an outlier with no significant flux excursion. The third row corresponds to a time instance that is both identified as an outlier and shows a large flux excursion.  Both of these latter images are poorly fitted by even 20 eigenimages and have been identified as outliers by their $R_{nl}$ values.}
\label{fig:outliers2}
\end{figure*}

There are many ways of using PCA to identify outliers in a set of images. A common method measures the Euclidean distance of each image in the hyperspace spanned by the set of eigenimages (often related to the Mahalanobis distance, \citealt{mahalanobis1936}). In implementations of outlier detection based on the Mahalanobis distance, the ensemble of images (or other data) is often standardized such that the distribution of pixel brightness within each image has been mean centered and scaled by its standard deviation. Because we have chosen not to standardize our data set, applying the Euclidean distance method directly to our PCA implementation would identify as outliers images that are simply brighter than the average image but without necessarily any substantial structural difference. In the context of using PCA to describe simulations of accreting black holes, we can easily identify such bright events by simply looking at large excursions of the total flux from the mean value. Our goal, instead, is to identify as outliers those snapshots with structures in the images that are substantially different from those of the typical snapshots. For this reason, we define a Euclidean distance using the fractional contribution of each eigenimage to the reconstruction of an image in the ensemble (see eq.~[\ref{eq:aprime}]).

We will consider a given snapshot (${\bf I}_n$) as typical, if it can be adequately reconstructed by the first $l$ eigenimages ($l$ can be chosen based on the particular distribution and application). To quantify the degree to which a snapshot is atypical, we define the quantity 
\begin{equation}
\label{eq:Qnl}
R_{nl} \equiv \frac{1}{\sum^m_{k=l+1}w_k}
\sum^m_{k=l+1} \left(\frac{a'_{nk}-\langle a'_{k}\rangle}{\sigma^2_{a_k}}\right)^2w_k\;
\end{equation}
that measures the weighted squared Euclidean distance from the mean of the distribution of $a'_{nk}$, scaled by the standard deviation of the distribution, and projected onto the subspace of eigenimages that is not being used in the reconstruction. Here, the mean of the distribution
\begin{equation}
\langle a'_{k}\rangle = \frac{1}{m}\sum_{n=1}^m a'_{nk}, 
\end{equation}
provides a measure of the average coefficient of an eigenimage needed to reconstruct the snapshots in a given set of images,
\begin{equation}
\sigma^2_{a'_{k}} = \frac{1}{m}\sum_{n=1}^m(a'_{nk}-\langle a'_{nk}\rangle)^2,
\end{equation}
shows the spread in that distribution, and $w_k$ is an appropriate weight function. We set the weight function to $w_k=\lambda_k$ because, in our implementation, the typical contribution of each eigenimage to any reconstruction is proportional to $\sqrt{\lambda_k}$ and we want to give more weight to the most dominant eigenimages when identifying outliers.

\begin{figure*}[t!]
\centering
\includegraphics{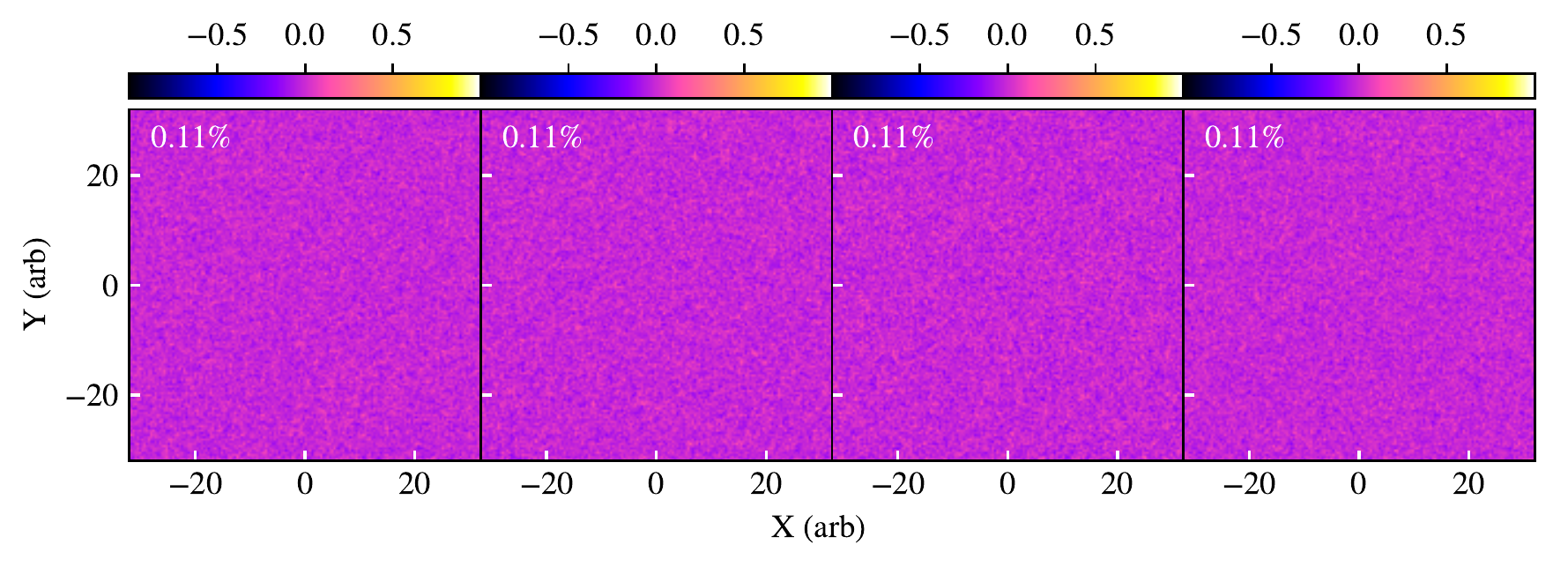}
\caption{The first 4 components of the PCA decomposition of 1024 images with purely Gaussian noise and their respective eigenvalues. The brightness of each pixel in these images is a random number taken from a Gaussian distribution centered at zero with a width of $\sigma = 0.5$.  Because each image is uncorrelated from the rest, all principal components have very similar eigenvalues and dimensionality reduction using PCA is not possible for this system.}
\label{fig:comp_gauss}
\end{figure*}

In Figure \ref{fig:outliers1} we show $R_{nl}$ for $l=5$, 10, and 20 for all snapshots in Model B as well as the normalized lightcurve. A number of time instances can be easily identified as atypical, i.e., with $R_{nl}\gg 1$, but these instances do not necessarily correlate with large brightness excursions. To examine this further, we show in Figure \ref{fig:outliers2} three original snapshots as well as the reconstructions using the first 5, 10, and 20 principal components. The top row (denoted as ``row 1" in Figure \ref{fig:outliers1}) shows an example of a time instance that is not identified as an outlier but corresponds to the largest flux excursion in this simulation. Clearly, this snapshot can be easily reconstructed by the first 10 eigenimages and has a low $R_{nl}$ value for all three values of $l$. This snapshot, despite being substantially brighter than the others, does not correspond to a significant structural change in the image. The second row (``row 2") shows a time instance that is identified as an outlier but does not correspond to a significant flux excursion. The morphology of the image is quite unusual compared to the rest of the simulation and a reconstruction with 20 eigenimages fails to capture the general structure of the image. The third row (``row 3") shows a time instance that is both identified as an outlier and  shows a significant flux excursion. The reconstruction of this snapshot with 20 eigenimages is also inadequate. 

These results demonstrate that, in our simulations, flux excursions and unusual image morphologies are not necessarily coincident but the two can be disentangled with the use of the quantity $R_{nl}$ that we have introduced here.

\section{Understanding the eigenvalue spectrum of PCA}\label{sec:results3}
\begin{figure}[t!]
\centering
\includegraphics{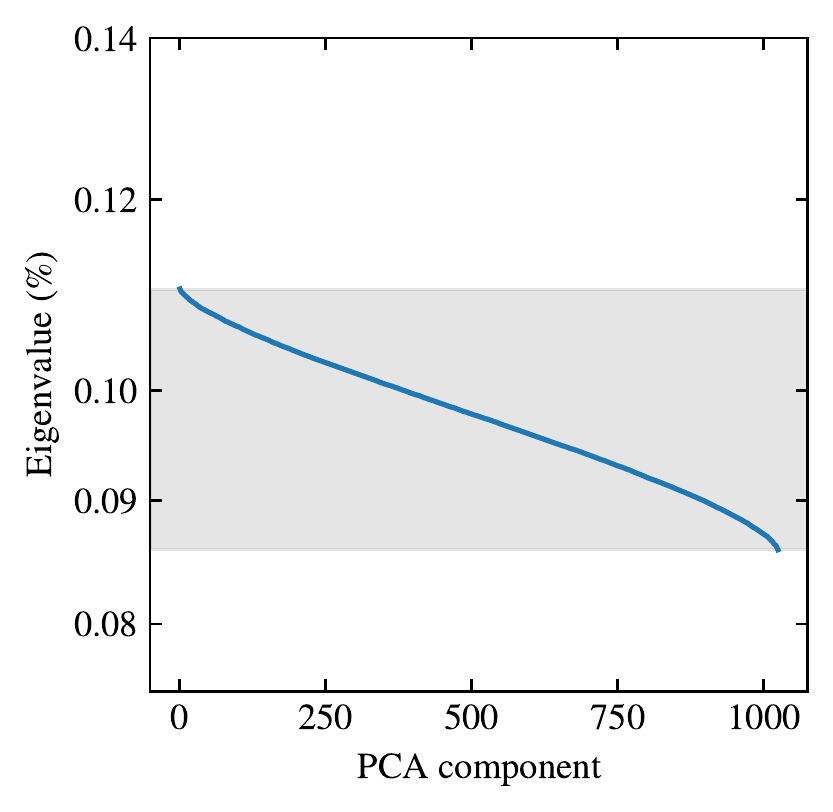}
\caption{The eigenvalue spectrum for the PCA decomposition of an ensemble of images with Gaussian noise. The gray rectangle represents the expected range of eigenvalues, given the statistical nature of Gaussian noise (see text). Note the very small range of the y-axis.}
\label{fig:spec_gauss}
\end{figure}

In this section we turn our attention to understanding the behavior of the spectrum of eigenvalues of the PCA decomposition of the GRMHD simulations. Specifically, we  focus on the higher-order components, which have small eigenvalues and contribute primarily to the small-scale, variable structures seen in the images. We aim to understand the origin of their eigenvalue spectra, which have the intriguing property of being power laws with very similar slopes in all simulations. This allows us to explore whether the spectra of eigenvalues are related to the underlying properties of GRMHD turbulence and, hence, whether measuring them in observations can help us better understand turbulence in accretion flows. 

The power-law shapes of the eigenvalue spectra are reminiscent of noise processes. For this reason, we begin by exploring the PCA eigenvalue spectrum of Gaussian noise in an image and then continue with a red-noise process.

\subsection{Gaussian Noise}

We consider a Gaussian noise model where the brightness of each pixel is a random number taken from a Gaussian distribution centered at zero. We perform PCA on 1024 images with independent realizations of Gaussian noise with a standard deviation of $\sigma=0.5$ over $512\times 512$ pixels. Figure \ref{fig:comp_gauss} shows the first few principal components and their respective eigenvalues. Any two images in the ensemble are statistically uncorrelated. However, the elements of matrix $L$ are not zero because small, non-zero residual correlations between any two images remain because of the finite number of pixels in each image and the statistical nature of noise. The eigenvalues of all components are similar, indicating that all of the principal components are of similar importance and dimensionality reduction is not possible for this configuration.

The presence of minor correlations between pairs of images leads to a distribution of eigenvalues of finite width. Because, in PCA, we count the eigenvectors in decreasing order of their eigenvalues, this distribution leads to a spectrum of eigenvalues with a non-zero slope. Figure \ref{fig:spec_gauss} shows the spectrum of the eigenvalues of our realization of the Gaussian noise model. The eigenvalues are normalized such that they sum to unity (see eq.~[\ref{eq:suml}]). Given that our simulation of Gaussian noise involves $m=1024$ images, there are 1024 non-trivial eigenvalues of similar magnitude with a mean of $1/1024\simeq 0.098\%$. To estimate the standard deviation of the distribution of eigenvalues, we consider the fact that there are $m N^2$ individual realizations of the Gaussian noise in the ensemble of $m=1024$ images with $N^2=512^2$ pixels each. We, therefore, expect the standard deviation of eigenvalues to be comparable to
\begin{equation}
\sigma = {1\over \sqrt{m N^2}}\simeq 0.06 \left(\frac{m}{1024}\right)^{-1/2}
\left(\frac{N}{512}\right)^{-1}\%.
\end{equation}
The full range of eigenvalues in our particular realization of images is from 0.0859\% to 0.1101\%, which corresponds to a width of $\approx 4\sigma$. In Figure \ref{fig:spec_gauss}, we show the range of $4\sigma$ around the expected mean magnitude of the eigenvalues to visualize this result.

In contrast to Gaussian noise, the spectra of PCA eigenvalues for our GRMHD simulations, including the power-law tails at large eigenvector numbers, do not depend on either the number of images or the number of pixels. We tested this by decreasing our spatial and temporal resolution by factors of 2 and 4 but preserving the total time span and image size. This behavior indicates that the structures present in our simulations are much larger than the pixel size; as a result, changing the number of pixels does not alter the PCA decomposition. A similar argument is valid for the lack of dependence on the number of images. For this reason, we now turn our attention to noise spectra with maximum power at scales that are larger than the pixel sizes.

\begin{figure}[t!]\centering
\includegraphics{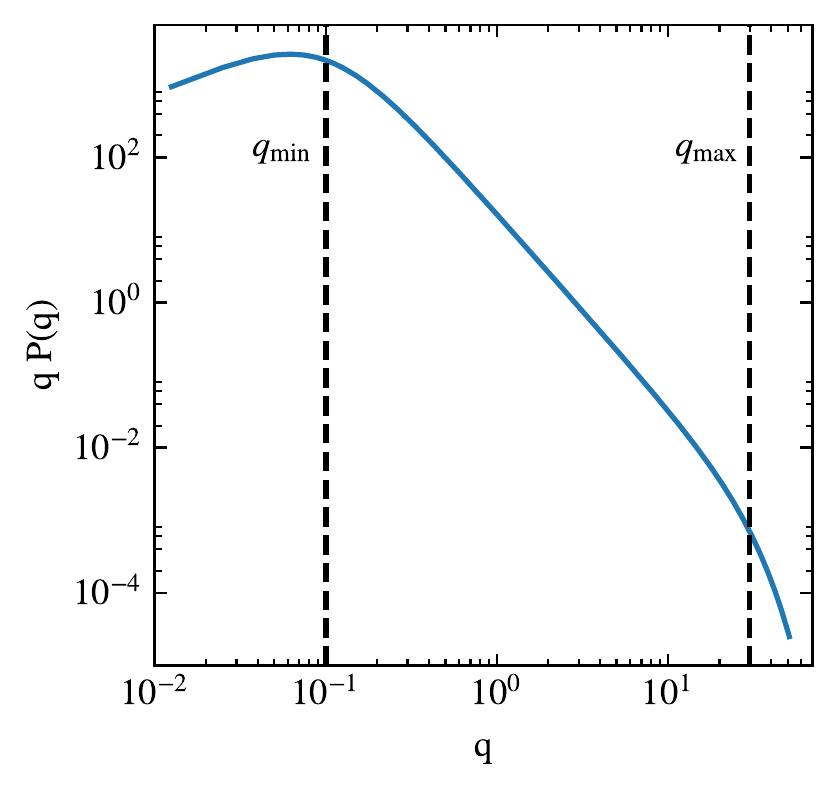}
\caption{The red-noise spectrum given in equation (\ref{eq:K_spec}) with $q_{\mathrm{max}}=30$, $q_{\mathrm{min}}=0.5$, and $\alpha=5/3$. The parameters $q_{\mathrm{min}}$ and $q_{\mathrm{max}}$ determine the locations of the first and second break in the spectrum, respectively, and $\alpha$ specifies the slope of the region between $q_{\mathrm{min}}$ and $q_{\mathrm{max}}$.}
\label{fig:K_spectrum}
\end{figure}

\begin{figure*}[t!]\centering
\includegraphics{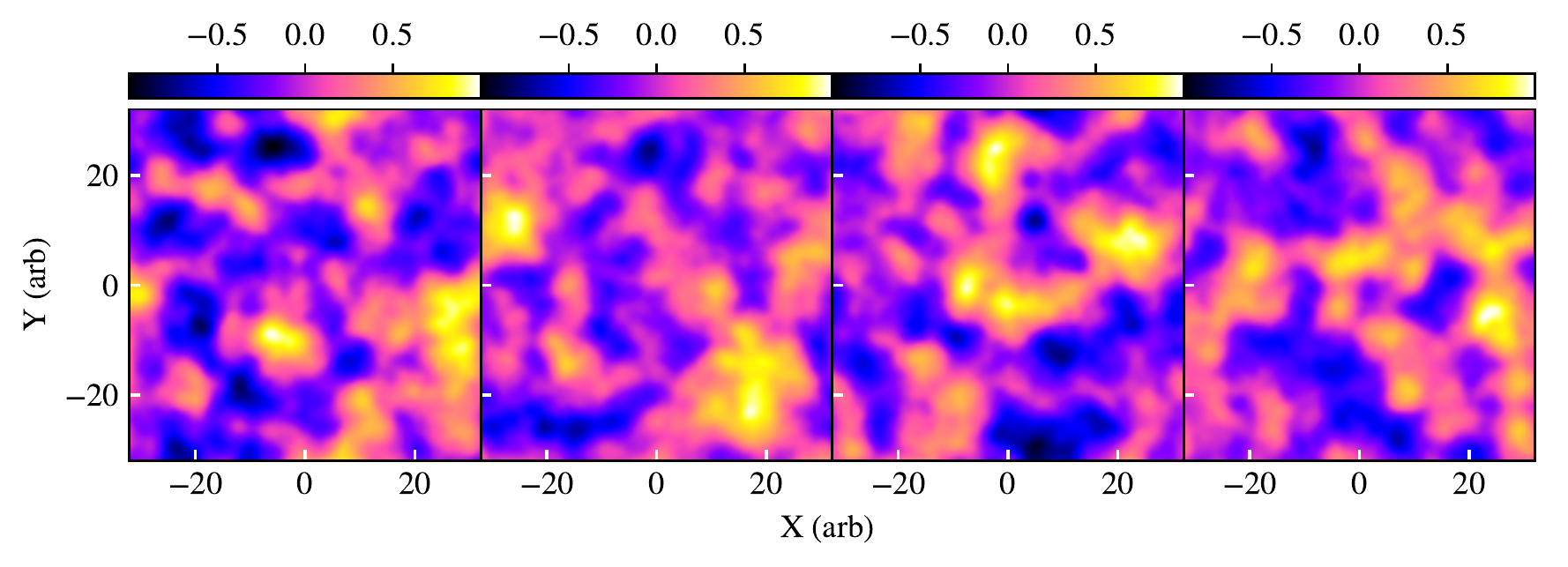}
\caption{Examples of images with different realizations of red-noise with the isotropic power spectrum shown in Figure \ref{fig:K_spectrum} and random phase fluctuations. As expected for $\alpha=5/3$, most of the power is at scales $\simeq 1/q_{\mathrm{min}}$. }
\label{fig:K_snaps}
\end{figure*}

\begin{figure}[t!]\centering
\includegraphics{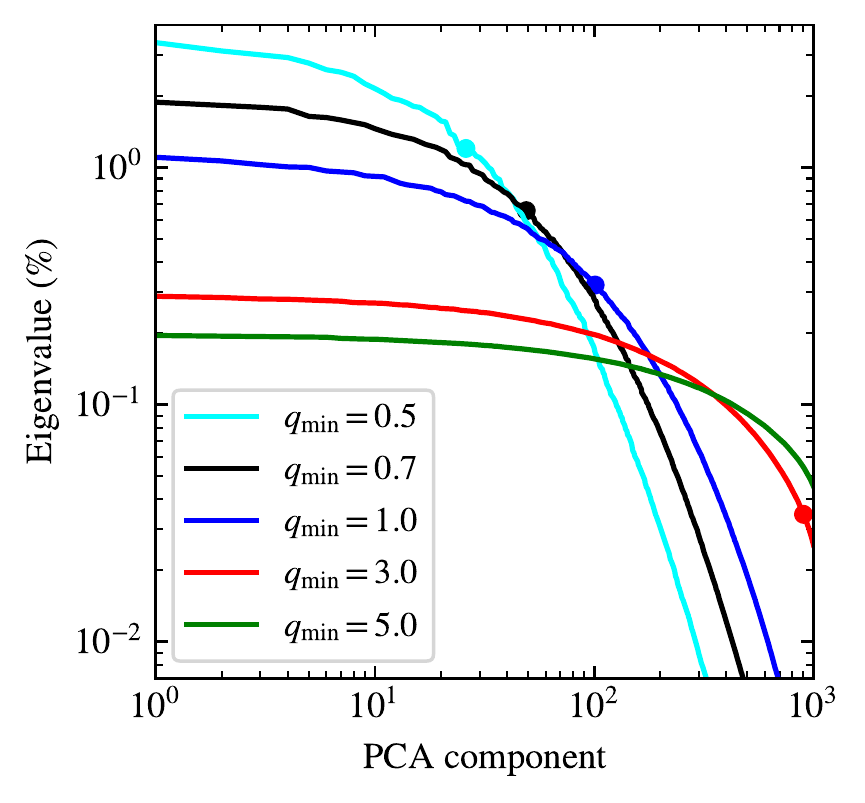}
\caption{The spectra of PCA eigenvalues for ensembles of images with isotropic red noise and for different values of the parameter $q_{\mathrm{min}}$; the remaining parameters are the same as in Figure~\ref{fig:K_spectrum}. For lower values of $q_{\mathrm{min}}$ the dominant scale of the structures in the images is larger and fewer PCA components are required to reproduce the majority of structure in the images. The filled circles on each curve indicate the number of PCA components that is equal to the approximate number of different noise structures that can fit in the image, i.e., where the number of components is equal to $L^2q_{\mathrm{min}}^2$, where $L$ is the size of the image.  }
\label{fig:K_spec_qmin}
\end{figure}

\begin{figure}[t!]\centering
\includegraphics{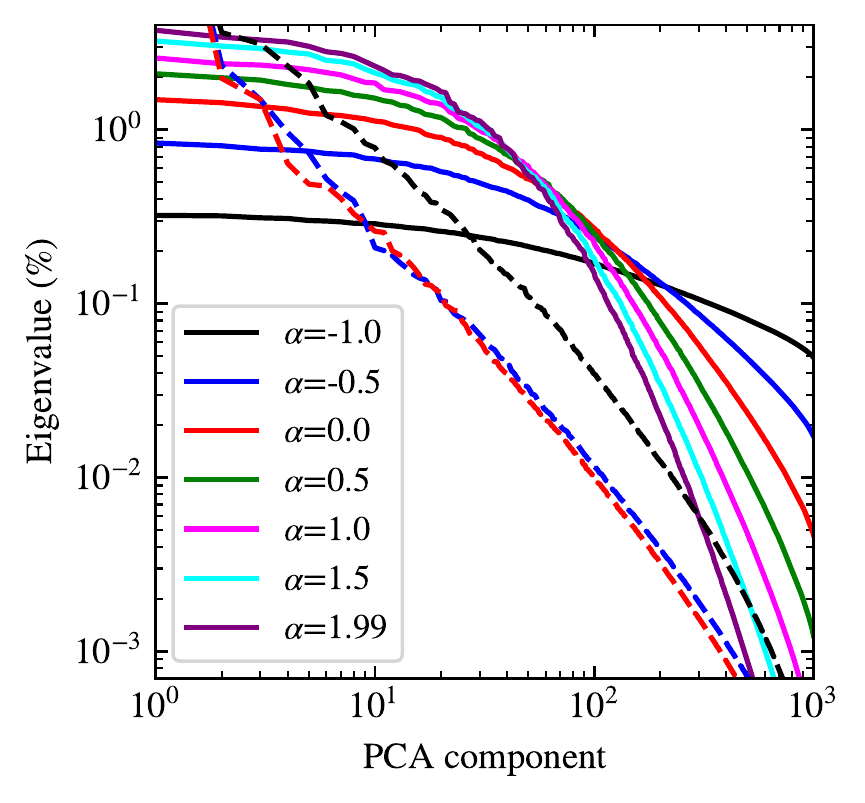}
\caption{The spectra of PCA eigenvalues for ensembles of images with isotropic red noise and for different values of the power-law index $\alpha$; the remaining parameters are the same as in Figure~\ref{fig:K_spectrum}.  For comparison, the eigenvalue spectra of the GRMHD models B, C, and D are also included in the black, blue, and red dashed lines respectively. The power law slope of the eigenvalue spectrum after the break depends strongly on $\alpha$.
\label{fig:alpha}}
\end{figure}

\begin{figure}[t!]\centering
\includegraphics{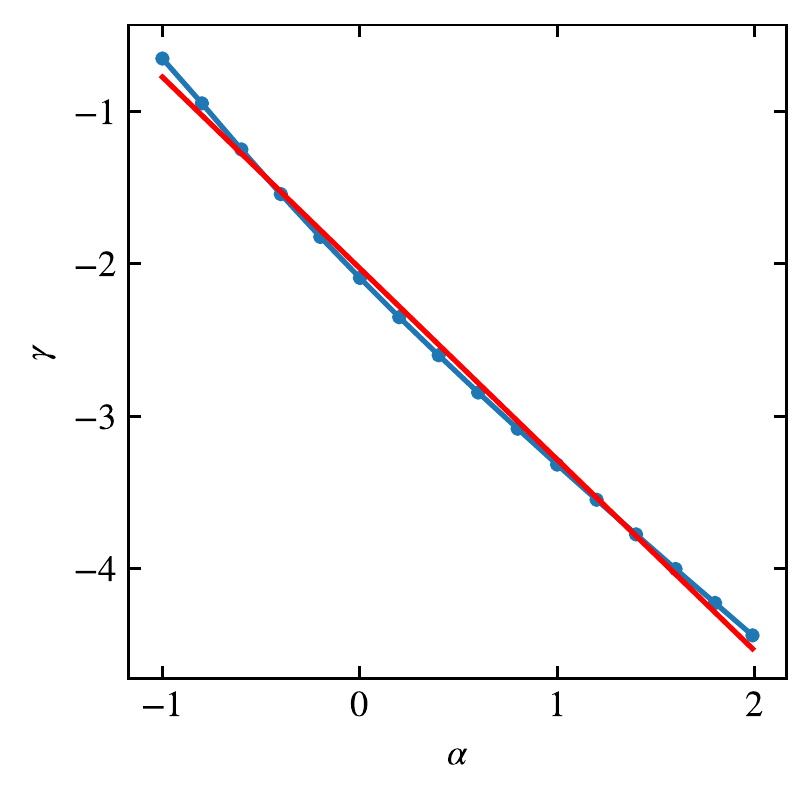}
\caption{The power-law index $\gamma$ of the spectra of PCA eigenvalues as a function of the power-law index $\alpha$ of the red-noise Fourier spectra that was used to generate the images. The blue circles are the values obtained numerically from fitting the eigenvalue spectra with power-law functions; the red line is a linear fit to the blue circles.}
\label{fig:slopeValpha}
\end{figure}

\subsection{Red Noise}

The spatial and time variability of images of accretion flows, such as those from \sgra, are expected to be approximated by red-noise power spectra. This is based both on the observationally measured flux variability of \sgra\ (\citealt{2008ApJ...688L..17M}, \citealt{2014MNRAS.442.2797D}) as well as on theoretical models (e.g., \citealt{2012ApJ...746L..10D}  and  \citealt{chan2015}). Other physical phenomena that affect black-hole images, such as refractive scattering in the intervening medium, are also expected to introduce noise at different characteristic scales (e.g., \citealt{2016ApJ...826..170J}). Because of such considerations, we consider here an ensemble of images with structure described by an isotropic, red-noise 2D Fourier spectrum given by
\begin{equation}\label{eq:K_spec}
P(q)= 2^{\alpha}\pi\alpha e^{-(q/q_{\mathrm{max}})^2}(q^2+q_{\mathrm{min}}^2)^{-(1+\alpha/2)}
\end{equation}
such that the image brightness at a location given by the transverse vector $\vec{r}$ on the image plane is
\begin{equation}
I(\vec{r})=I_0\int d^2q P(q)\exp\left[-i \vec{q}\cdot\vec{r}\right]\;.
\end{equation}
Here, $q_{\mathrm{min}}$ and $q_{\rm max}$ determine the location of the first and second breaks in the spectrum and consequently the sizes of the largest and smallest structures in the images, respectively. The parameter $\alpha$ determines the slope of the power spectrum in the region between $q_{\mathrm{min}}$ and $q_{\mathrm{max}}$. 

Figure \ref{fig:K_spectrum} shows a plot of this spectrum with $q_{\mathrm{max}}=30$, $q_{\mathrm{min}}=0.5$, and $\alpha=5/3$. As expected, for $\alpha=5/3$, there is very little power at the small scales where $q_{\mathrm{max}}$ is relevant; the majority of the power is at the larger scales related to $q_{\mathrm{min}}$. Figure \ref{fig:K_snaps} shows some examples of images with different red-noise realizations. By construction, the structures in these images are almost entirely resolved in an $N \times N$ image, as long as the size of each pixel is much smaller than $1/q_{\rm max}$. 

In order to investigate the effect of red noise on the PCA of images, we construct numerous sets of 1024 images for different values of the red noise parameters, such as those in Figure \ref{fig:K_snaps}, and perform PCA on the set. We now explore the dependence of the PCA decomposition of these images on the parameters of the red-noise spectrum.

The spectrum of PCA eigenvalues of red noise does not depend on the number of pixels $N$ per image, as long as the size of the dominant scale of the noise is fully resolved, i.e., as long as $L/N<<1/q_{\mathrm{min}}$, where $L$ is the linear size of the image. This is similar to the PCA results for the images of the GRMHD simulations and unlike those of the Gaussian noise simulations discussed earlier. The eigenvalue spectrum is also independent of $q_{\mathrm{max}}$ as long as $q_{\mathrm{max}}>>q_{\mathrm{min}}$ and $\alpha>-1$ because, if these conditions are met, there is negligible power at scales $\sim 1/q_{\mathrm{max}}$ to affect the PCA decomposition significantly.

Figure \ref{fig:K_spec_qmin} shows the spectrum of PCA eigenvalues and its dependence on $q_{\mathrm{min}}$. For $\alpha>-1$, $1/q_{\mathrm{min}}$ determines the size of the dominant scale of the noise structures. The number
of dominant noise structures that can fit in an image of size $L$ is
\begin{equation}
n= \left(L\over1/q_{\mathrm{min}}\right)^2=L^2q_{\mathrm{min}}^2\;.
\label{eq:n_comp}
\end{equation}
We, therefore, expect, following the discussion in \S3, that this number corresponds to the number of dominant PCA components. This is shown in Figure~\ref{fig:K_spec_qmin}, where the filled circle on each spectrum corresponds to the eigenvalue of the $n-$th PCA component given by the above relation. Clearly, as $q_{\rm min}$ increases, the structures on the images become smaller and more PCA components are necessary to reconstruct with fidelity the original ensemble of images.

For a given value of the parameter $q_{\mathrm{min}}$, the number of images $m$ in the ensemble determines whether the spectrum of eigenvalues has converged or not. Indeed, as we discussed above, for small values of $q_{\rm min}$, which correspond to large dominant noise structures, a small number of eigenimages is required to reconstruct with fidelity the full ensemble of images. In this case, i.e., as long as $m\gg n= L^2q_{\mathrm{min}}^2$, the eigenvalue spectrum has converged and its shape depends only very weakly on the number $m$ of images in the ensemble.

Figure~\ref{fig:alpha} shows the spectrum of PCA eigenvalues for images with Fourier spectra characterized by different power-law indices $\alpha>-1$. As in Figure~\ref{fig:K_spec_qmin}, the spectra are relatively flat until the $n-$th PCA component but then turn into power laws with indices that appear to be correlated with $\alpha$.  For $\alpha\lesssim -1$, which we do not show, the dominant noise structures occur at the small scales $\simeq 1/q_{\rm max}$ and the resulting eigenvalue spectra are flat with very weak dependence on $\alpha$. 

We further explore the dependence of the eigenvalue spectra on $\alpha>-1$ by fitting the higher components of each eigenvalue spectrum with a power-law function of the form $\lambda_k\sim k^{-\gamma}$ and show in Figure~\ref{fig:slopeValpha} the dependence of the fitted power-law index $\gamma$ on $\alpha$. We find this dependence to be
\begin{equation}
\gamma\simeq\frac{5}{4}\alpha+2\;.
\label{eq:gamma_vs_alpha}
\end{equation}
Note that, for the simulations used in generating Figure~\ref{fig:slopeValpha}, we set $q_{\mathrm{min}} = 0.1$ to force the breaks of the eigenvalue spectra to occur at low PCA components and, thus, to allow for a more accurate determination of the power-law index of the spectra. This result demonstrates that the 2D Fourier spectrum of the structures in the image plane determine in a predictable way the high-end spectrum of PCA eigenvalues and, therefore, the latter can be used to infer the former.

\subsection{The Small-Scale Structures of Black-Hole Images from GRMHD Simulations}

We compare in Figure~\ref{fig:alpha} the spectra of PCA eigenvalues from the black-hole images of GRMHD simulations to those of the images with red-noise Fourier spectra. The large range of eigenvalues in the GRMHD simulations is clearly inconsistent with the small expected range of eigenvalues for purely Gaussian noise (see also Fig.~\ref{fig:spec_gauss}). This suggests a more complex origin of image structure and variability than what has been assumed in the past (cf.~\citealt{2016ApJ...820..137B}). 

The spectra of PCA eigenvalues for the images of GRMHD simulations become power laws after only the first handful of PCA components. This suggests that $L q_{\rm min}$ for these simulations is a small number (see eq.~[\ref{eq:n_comp}] and Figure~\ref{fig:K_spec_qmin}) and, therefore, that the typical scale of variable structure in the images is comparable to the size of the images themselves. In other words, it is comparable to the size of the black-hole shadow. This is consistent with the discussion in~\citet{2017ApJ...844...35M, Medeiros2018}, who attributed the variability of the simulated interferometric amplitudes and closure phases to overall changes in the widths of the crescent-like images as well as to the appearance and disappearance from the images of large, hot, and, therefore, bright magnetic flux tubes that orbit the black hole.

The power-law indices in the eigenvalue spectra of the images from GRMHD simulations are nearly independent of the underlying model and equal to $\gamma\simeq 1.3$. Using equation~(\ref{eq:gamma_vs_alpha}), we find that this implies a power-law index for
the 2D Fourier spectrum of the variable structures of $\alpha\simeq -0.5$. It is important to emphasize here that this power-law index characterizes the 2D Fourier spectrum of the variable structures, which are determined in a complex, non-linear way by the anisotropies in the density, temperature, emissivity, magnetic field, and lensing in the vicinity of the black hole. It is, therefore, not surprising that the inferred value of $\alpha$ does not reflect the underlying power-spectrum of the MHD turbulence in the accretion flow.

\section{Conclusion}\label{sec:conc}

Understanding the horizon-scale millimeter emission around an accreting black hole
requires a two-pronged approach.  One component is to utilize our best understanding of the physics
to generate high-fidelity GRMHD simulations of the morphology of the emission. The second
is to use an interferometer, such as the EHT, to test the understanding of the physics with real observations.  In both components, there is a common theme: the question of how we characterize and extract the salient information in an ensemble of images. In this paper, we have demonstrated that PCA offers an effective tool for this task over a wealth of different problems.

Focusing purely on the simulations, we showed that PCA offers an extremely compact representation of the theoretical millimeter images.  Each simulation comprises over 1000 distinct images, yet we find that we can represent {\it most} of the images with only a few to a few dozen eigenimages, depending on the desired fidelity.  Moreover, recognizing
images poorly represented by the leading eigenimages is critical and represents another useful application of PCA.   As detailed in the Introduction, temporal variability of the strength and morphology of the millimeter emission close to the horizon is a phenomenon that can limit or compromise the construction of interferometric images.  Knowledge of the amplitudes of the eigenimages needed to represent any given image can be used to define a simple scalar metric
that flags outliers in either the simulations or observations.  This approach has already provided the realization that outliers may be more subtle than had been presumed.  It had been supposed that flares in flux would correspond to events in which the emission morphology would show strong departures from the average form. Yet the outlier metric $R_{nl}$ (see Eq.~\ref{eq:Qnl}) allowed us to identify both images that had unusual morphology with no significant excursion in flux, as well as flare events that had perfectly ordinary morphology.  As useful as this particular metric is in this work, however, we emphasize that other metrics and classifiers can be constructed from the locations of the simulated images in their eigenspace.  Our goal here is not to strongly advocate any particular metric but to provide a useful example of what is possible within the PCA representation.

Apart from the identification of outliers, we also demonstrated the use of the eigenvalue spectrum to characterize the properties of the noise and turbulent structure in the simulations.   This approach shows a path for allowing the rapid quantitative evaluation of GRMHD simulations over a significant timespan of accretion.  As with the outlier metric $R_{nl}$, other diagnostic metrics can be built around the locations or trajectories of the simulated images as a function of time in their eigenspace.

Lastly, we showed that PCA may be applied directly to the analysis of interferometric data because the Fourier transform of the principal components of a set of images is equivalent to the principal components of the set of Fourier transformed images. Coupled with the dimensionality reduction that we discussed above, this property opens the possibility of using PCA for efficient image reconstruction from sparse interferometric data. In parallel,  the PCA approach can be incorporated into the Bayesian inference method discussed in \citet{2016ApJ...832..156K}, in order to generate efficient comparisons of EHT data to large suites of GRMHD simulations. We will explore these avenues in future work.

\acknowledgements

\acknowledgements
L.\;M.\ acknowledges support from NSF GRFP grant DGE~1144085.
D.\;P.\ and F.\;O.\ gratefully acknowledge support from NSF PIRE grant 1743747, NSF AST-1715061, and Chandra Award No. TM8-19008X for this work. All ray tracing and PCA calculations were performed with the \texttt{El~Gato}
GPU cluster at the University of Arizona that is funded by NSF award
1228509. We thank Dan Marrone for many useful discussions and comments on this manuscript.


\bibliography{main,my}
\end{document}